\documentclass[amssymb,amsmath,useAMS,prd,aps,twocolumn,superscriptaddress,preprintnumbers,nofootinbib]{revtex4}
\usepackage{pslatex}
\usepackage[pdftex]{graphicx}
\usepackage{psfrag}
\usepackage{epsfig}
\usepackage{color}
\usepackage{cancel}
\usepackage{slashed}
\usepackage{ulem}

\def\refe@jnl#1{{#1}}
\def\aj{\refe@jnl{Astron.~J.}}
\def\araa{\refe@jnl{Annu.~Rev.~Astron.~Astrophys.}}
\def\apj{\refe@jnl{Astrophys.~J.}}
\def\apjl{\refe@jnl{Astrophys.~J.~Lett.}}
\def\aap{\refe@jnl{Astron.~Astrophys.}}
\def\mnras{\refe@jnl{Mon.~Not.~R.~Astron.~Soc.}}
\def\prd{\refe@jnl{Phys.~Rev.~D}}
\def\fcp{\refe@jnl{Fund.~Cos.~Phys.}}
\def\physrep{\refe@jnl{Phys.~Rep.}}
\def\physlett{\refe@jnl{Phys.~Lett.}}

\def\invisible#1{  }

\def\lsim{\mathrel{\lower4pt\hbox{$\sim$}} 
\hskip-9.5pt\raise1.6pt\hbox{$<$}\;} 
 
\def\gsim{\mathrel{\lower4pt\hbox{$\sim$}} 
\hskip-9.5pt\raise1.6pt\hbox{$>$}\;}

\begin{document}
\preprint{ULB-TH/15-04}

\title{Inert Scalar Doublet Asymmetry as Origin of Dark Matter}

\author{Mika\"el Dhen and Thomas Hambye}
\email{mikadhen@ulb.ac.be;thambye@ulb.ac.be}
\affiliation{Service de Physique Th\'eorique\\
 Universit\'e Libre de Bruxelles\\ 
Boulevard du Triomphe, CP225, 1050 Brussels, Belgium}


\begin{abstract}
In the inert scalar doublet framework, we analyze what would be the effect of a $B-L$ asymmetry that could have been produced in the Universe thermal bath at high temperature. We show that, unless the ``$\lambda_5$" scalar interaction is tiny, this asymmetry is automatically reprocessed in part into an inert scalar asymmetry that could be at the origin of   dark matter today. Along this scenario, the inert mass scale lies in the few-TeV range and direct detection constraints require that the inert scalar particles decay into a lighter dark matter particle which, as the inert doublet, is odd under a $Z_2$ symmetry.
\end{abstract}

 \maketitle
 
 \section{Introduction}
 
 The similarity of baryonic and dark matter (DM) abundances, determined by the observation of the Cosmic Microwave Background (CMB) anisotropies, $\Omega_{DM}/\Omega_B=5.4\pm 0.1$ \cite{Planck:2015xua}, has motivated a long series of scenarios were both abundances have a related or even very same origin. Since the baryon asymmetry is to a very good approximation totally asymmetric -- no primordial population of antibaryons has been observed in the Universe -- a common origin of both abundances suggests that the DM abundance today would be associated to the generation of a DM particle-antiparticle asymmetry (see e.g.~the reviews of Refs.~\cite{Davoudiasl:2012uw,Petraki:2013wwa,Zurek:2013wia,Boucenna:2013wba}). 
In the following, we will show   how this can be realized from the generation of a scalar inert doublet asymmetry. The 
inert scalar doublet DM framework (IDM) \cite{Ma:2006km,Barbieri:2006dq,LopezHonorez:2006gr,Hambye:2009pw} simply  consists in adding to the Standard Model (SM) a single scalar doublet, $H_2$, odd under a $Z_2$ symmetry. The most general scalar potential is in this case  
\begin{align}
V&=   m_1^2|H_1|^2+  m_2^2|H_2|^2+   \lambda_1|H_1|^4+ \lambda_2|H_2|^4 \nonumber \\
&+ \lambda_3|H_1|^2|H_2|^2+ \lambda_4|H^\dagger_1 H_2|^2+ \frac{\lambda_5}{2}\left[\left(H_1^\dagger H_2\right)^2+h.c.\right]\,, \label{IDMpot}
\end{align}
where the SM and the inert scalar doublets can be written as
\begin{equation}
 H_1=\left(\begin{array}{c}
\phi^+\\
v/\sqrt{2}+\phi^0
\end{array}\right)\quad \text{and}\quad
 H_2=\left(\begin{array}{c}
\eta^+\\
\eta^0
\end{array}\right)\,,
\end{equation}
with  $\phi^0\equiv(h+i\phi^3)/\sqrt{2}$ and  $\eta^0\equiv  (H^0+i A^0)/\sqrt{2}$.
In the scalar potential, $m_2^2$ is assumed to be positive to insure that $H_2$ doesn't acquire a vev, so that it's lightest (neutral) component is stable, unless there exists a lighter $Z_2$ odd particle into which it can decay.    This is the possibility we  will ultimately  consider in the following, in order to satisfy the direct detection constraints, see Section~\ref{sec:H2-decay}. But before, up to the end of Section III, let's restrict ourselves  to the usual IDM setup and see what happens in this minimal framework.

 Prior to  electroweak symmetry breaking (EWSB), all $H_2$ components have mass $m_{H_2}=m_2$, whereas  after EWSB ($v=246$~GeV), they get split in mass 
\begin{align}\label{mass-splitting}
 		m^2_{H^0}=m^2_2+\lambda_{H^0} v^2\, , \ m^2_{A^0}=m_2^2+\lambda_{A^0} v^2\, , \ m^2_{\eta^+}=m_2^2+\lambda_{H^c} v^2 \, ,
\end{align}
with $\lambda_{H^c}=\lambda_3/2$ and $\lambda_{H^0,A^0}=\left(\lambda_3+\lambda_4\pm\lambda_5\right)/2$. 
In the following, we will assume, without loss of generality, that $\lambda_5$ is negative, so that $H^0$ is lighter than $A^0$. 
Various well-known constraints hold on the parameters of the theory. Tree level vacuum stability requires $\lambda_{1,2}>0$,  $\lambda_{H^0,A^0,H^c}>-\sqrt{\lambda_{1}\lambda_{2}}\approx -0.36 \sqrt{\lambda_{2}}$. EW precision test observables require
$\Delta T\simeq \left(m_{\eta^+}-m_{A^0}\right)\left(m_{\eta^+}-m_{H^0}\right)/  12 \pi^2\alpha v^2\lesssim  10^{-1}$. $Z$ decay width constraint at LEP requires $m_{A^0} + m_{H^0} > m_Z$ and $m_{\eta^+}>m_Z/2$. 
Direct detection constraint importantly requires that the $Z$ exchange diagram is kinematically forbidden, i.e.~$ m_{A^0}-m_{H^0}\gtrsim   \mu_r \beta_{DM}^2/2$, where $\beta_{DM}c$ is the DM halo velocity with respect to the earth, and $\mu_r=m_{H^0}m_{{\cal N}}/(m_{H^0}+m_{{\cal N}})$ is the reduced mass of the system for the nucleus ${\cal N}$ used by the experiment. For $m_{H^0}\gg m_{\cal N}$ and Xenon nucleus, using an average velocity of $\sim 270$~km/s, this constraint can be rephrased   as $ m_{A^0}-m_{H^0}\gtrsim \delta m_{min}\sim50$~keV.
Taking into account the velocity distribution around this central value,
and the recoil energy sensitivity of the experiments, the minimum splitting becomes  $\delta m_{min}\sim180$~keV, although 
a more robust constraint is  $\delta m_{min}\sim   100$~keV \cite{Nagata:2014aoa},  which translates as
\begin{equation}\label{direct-detection}
|\lambda_5| \gtrsim 3.3\cdot 10^{-6}\cdot \left(\frac{m_{H^0}}{\text{TeV}}\right) \cdot \left(\frac{\delta m_{min}}{100\,\hbox{keV}}\right)\, .
\end{equation}
It is well known that the IDM can account for the observed DM relic abundance via the usual freeze-out mechanism, and be in agreement with direct detection constraints, for DM masses in the ranges $\sim [50,80]$~GeV and above $\sim 540$~GeV, up to the $\sim 40$-$50$~TeV unitarity bound \cite{Barbieri:2006dq,LopezHonorez:2006gr,Cirelli:2005uq,Hambye:2009pw,Fonseca:2015gva}. 
As we will show, it could also be responsible for the DM relic density in an asymmetric way.

 \section{Asymmetric production of the DM relic density}

As said above, in this Section and the following one we stick to the minimal IDM framework, as defined in the introduction. 
Let us make two simple starting assumptions. First, let us assume that the symmetric component of the relic density left after freeze-out is smaller than the observed value. Fast SM gauge scatterings automatically care for that for $m_{H^0}$ within the $\sim 120-540$ GeV range, whereas for other values of $m_{H^0}$ large enough $\lambda_{3,4}$ interactions can take care of that \cite{Hambye:2009pw}.  This implies a symmetric annihilation cross section larger than the usual thermal freeze-out value  $\sim 1$~pb, 
which means a freeze-out temperature $T_{fo}$ smaller than the usual $T_{fo}\sim m_{H^0}/25$ value.  Second, let us assume that a $B-L$ asymmetry has been generated at a temperature $T_{B-L}$ above  $m_{H_2}$ and above the  EWPT temperature $T_{EW}$ (which we take as the temperature where the vacuum expectation value of the SM scalar field becomes sizable, that is $T_{EW} \approx 165$~GeV from Ref.~\cite{D'Onofrio:2014kta}).  We do not care about the way this $B-L$ asymmetry could have been generated. It could be due for example to the straightforward leptogenesis mechanism. Note that, as well-known, if a $B-L$ asymmetry is generated at high temperature, a $H_1$-$H_1^*$ asymmetry will also be created automatically at high temperature from   thermal equilibrium SM interactions \cite{Harvey:1990qw}.

If a $B-L$ (and thus $H_1$ asymmetry) is created at high temperature, an inert doublet $H_2$-$H_2^*$ asymmetry is to be expected too. The scalar potential of Eq.~(\ref{IDMpot}) contains the $\lambda_5$ interaction which uniquely does not conserve the number of $H_2$ minus the number of $H_2^*$ (as well as the number of $H_1$ minus the number of $H_1^*$). This interaction is in thermal equilibrium at $T\sim m_{H_2}$ if at this temperature the corresponding $\Gamma_{\lambda_5}$ scattering rate, given in the Appendix, is larger than the Hubble rate, which gives the condition 
 \begin{equation}\label{lambda5-condition}
|\lambda_5|\gtrsim  10^{-6} \cdot \left( m_{H_2}/\text{TeV}\right)^{1/2}\ .
\end{equation}  
If Eq.~(\ref{lambda5-condition}) is satisfied, the $\lambda_5$ interaction equilibrates the $H_2$ and $H_1$ (and $B-L$) asymmetries.\footnote{Actually, in the few-TeV asymmetric inert DM scenario considered in Ref.~\cite{Arina:2011cu}, it is assumed instead that the $\lambda_5$ interaction could have never been in thermal equilibrium. In this case, the DM asymmetry would have been created explicitly at high energies, basically independently of the $B-L$ asymmetry.   If the inert scalar is the DM particle, it turns out that the lower bound of Eq.~(\ref{direct-detection}) implies that Eq.~(\ref{lambda5-condition})  
must anyway hold for TeV masses.
For instance, Eq.~(\ref{direct-detection}) with a $100$~keV (180~keV) mass splitting  implies Eq.~(\ref{lambda5-condition}) for  $m_{H_2}\gtrsim 100$~GeV (30~GeV). 
}
In particular even if, as we assume here, no $H_2$ asymmetry is created at high energies, such an asymmetry will be created anyway as soon as the $B-L$ asymmetry is created. 
In other words, the inert DM model contains an interaction which basically implies that  ``Higgsogenesis" \cite{Servant:2013uwa} production of a DM asymmetry is at work.\footnote{In Ref.~\cite{Servant:2013uwa}, a $X_1$ fermion singlet DM framework is considered with an extra $X_2$ fermion doublet and an $X$-symmetry. An $X_2$ asymmetry is created from a $X$-symmetry violating $X_2^2H_1^2$ non-renormalizable interaction, which is afterwards reprocessed into a $X_1$ symmetry through $X_2$ decays.  Asymmetric frameworks based on the equilibration of the SM scalar asymmetry with a dark sector asymmetry, based on several new dark sector particles, or based on various possibilities of a $SU(2)_L$ multiplet, can also be found in Refs.~\cite{Dulaney:2010dj} and \cite{Boucenna:2015haa} respectively.}  
Note that the scenario could work also the other way around, i.e.~a primordial DM asymmetry could be at the origin of baryogenesis via the same $\lambda_5$ equilibration interaction, a possibility we will not consider here (for a scenario of this kind see Ref.~\cite{Davidson:2013psa}).

In the following, we will consider in details and chronologically 
what happens when the temperature of the Universe cools  down from $T\gg m_{H_2}$  to today $T\ll T_{EW}$, crossing  $m_{H_2}>T_{\lambda_5}>T_{fo}>T_{EW}$, with $T_{\lambda_5}$ the temperature where the scattering induced by the $\lambda_5$ interaction decouples and $T_{fo}$ the freeze-out temperature at which the total annihilation cross section decouples.
Given that the inert doublet components undergo gauge interactions,  $T_{\lambda_5}$ is sizably larger than $T_{fo}$, unless $\lambda_5$ is of order one, which as we will see is not a viable option for the case we are interested in (where the DM asymmetry is responsible for most of the relic density). Similarly, as we will see, $T_{fo}\gtrsim T_{EW}$, i.e.~few-TeV DM, is also generically necessary in order to have a viable scenario (as in the scenario of Ref.~\cite{Arina:2011cu}), but some violation of this inequality is possible. A sketch of the scenario, applied to our framework,  is shown in Fig.~\ref{sketch}.

\subsection{$\mathbf{T\gtrsim m_{H_2}}$}

At temperature above $T_{EW}$, all 4 inert doublet components have a common mass  $m_{H_2}= m_2$. If Eq.~(\ref{lambda5-condition}) is satisfied, the chemical potential of both scalar doublets are equal,
$\mu_{H_2}=\mu_{H_1}$. Together with
the usual SM chemical equilibrium relations (from
 thermal equilibrium SM processes \cite{Harvey:1990qw}), the $\mu_{\eta^+}= \mu_{\eta^0}$ relation (from 
  e.g.~$\eta^+ \eta^{0*}\leftrightarrow SM$ 
  processes), and the hypercharge relation  
 \begin{equation}\label{hypercharge-condtion}
\sum_i\left(  \Delta _{Q_i}+4\Delta _{u_i}-2\Delta _{d_i}-\Delta _{\ell_i}-2\Delta _{e_i}\right)+  \Delta _{H_1} +  \Delta _{H_2}  =0 \ ,
\end{equation}
it simply gives   
 \begin{equation}
	\Delta _{H_2} = \Delta _{H_1}
	= -\frac{4}{23}\Delta _{B-L}\,.\label{eq:eq-cond-fast-mH2}
\end{equation}
From now on, we define  for each species $X$ the asymmetry $\Delta_X\equiv Y_X-Y_{\bar X}$ and the total density $\Sigma_X\equiv Y_X+Y_{\bar X}$, where $Y_X\equiv n_X/s$ is the particle number density-to-entropy ratio of $X$. Since we are dealing with asymmetries, we also define the number of degrees of freedom by summing the number of particles (or antiparticles but not both), i.e.~$g_X=1$ for a $SU(2)_L$ singlet, and $g_X=2$ for a doublet.

As well-known, for similar $B-L$ and DM asymmetries,  the DM relic density constraint requires $m_{DM}$ to have a mass of few GeV (more exactly, from Eq.~(\ref{eq:eq-cond-fast-mH2}) and taking into account the $Y_{B-L}$ to $Y_B$ ratio which holds in this case, Eq.~(\ref{B-below-EW-high-T}) below,  one would need $m_{DM}\approx 10$~GeV). As this possibility is excluded by collider constraints, this implies that a subsequent suppression of the DM asymmetry by a factor of $\sim (10\,~\hbox{GeV}/m_{DM})$ must necessarily occur. Two different types of suppressions can naturally take place. A first one is a Boltzmann suppression from asymmetry violating scatterings, used in several other DM models, see e.g.~Refs.~\cite{Nardi:2008ix,Servant:2013uwa}. In our scenario, it can arise from the $\lambda_5$ interaction within the period $m_{H_2}>T>T_{\lambda_5}$. The other possible suppression can arise later when $T\lesssim T_{EW}$ from the combined effect of DM oscillations and symmetric annihilations. 

\subsection{$\mathbf{m_{H_2}\gtrsim T\geq T_{\lambda_5}}$}

Once the temperature drops below $m_{H_2}$, if the $\lambda_5$ interaction goes on to be in thermal equilibrium, the $H_2$ asymmetry gets Boltzmann   suppressed. 
This  can be directly seen from the Boltzmann equation of  $\Delta_{H_2}$, valid for $T\geq T_{EW}$, 
\begin{equation}\label{boltzmann-eq}
\frac{d \Delta_{H_2} }{dz}=-\frac{4}{sHz}\ \Big( \frac{\Delta _{H_2}}{Y_{H_2}^{eq}} -\frac{\Delta _{H_1}}{Y_{H_1}^{eq}}    \Big)\ \gamma_{\lambda_5}	\ ,
\end{equation} 
where $z\equiv m_{H_2}/T$, $H(z)$ is the Hubble rate and   
$\gamma_{\lambda_5}(z)$ is the reaction density of the $\lambda_5$ scatterings, given in Appendix.    It includes both pair annihilation/creation $H_2 H_2 \leftrightarrow H_1 H_1$ and ``t-channel" $H_2 \bar{H}_1\leftrightarrow H_1 \bar{H}_2$ processes. 
The $\lambda_5$ interaction leaves intact the sum of the asymmetries of $H_1$ and $H_2$ but not each asymmetry individually. Once $T$ drops below $m_{H_2}$, the  first term in the r.h.s.~of Eq.~\eqref{boltzmann-eq} 
is enhanced with respect to the  second 
term by the fact that $Y_{H_2}^{eq}$ is Boltzmann suppressed, unlike $Y_{H_1}^{eq}$. This Boltzmann suppression of the asymmetry lasts until the $\lambda_5$ induced scatterings decouple, at $T=T_{\lambda_5}$, when $\Gamma_{\lambda_5}\simeq H$. Quantitatively, this can be accounted by the usual $k$-factor which gives
the asymmetry as a function of the temperature
\begin{eqnarray}
\label{YH2kfactor}
\Delta _{H_2}&=&\frac{T^2}{6s}\ g_{H_2}\ \mu_{H_2}\ k(z) \ ,\\
\label{k-factor}
 k\left(z\right)&\equiv& \frac{6}{4\pi^2} \int_0^\infty x^2 \sinh^{-2} \left(\frac{1}{2}\sqrt{x^2+z^2}\right)dx \ .
\end{eqnarray}
Down to $T_{\lambda_5}$, the chemical potential relation
$\mu_{H_2}=\mu_{H_1}$ still holds, and we get
 \begin{equation}\label{eq:eq-cond-fast-high-T}
	\Delta  _{H_2}(z) =\frac{k\left(z\right)}{2} \Delta  _{H_1} \, .
	\end{equation}
This is nothing but the solution which makes the r.h.s.~of Eq.~(\ref{boltzmann-eq}) to vanish.
Together with the other chemical potential relations above, the asymmetry reads at $T= T_{\lambda_5}$ (similarly to the fermion doublet case of Ref.~\cite{Servant:2013uwa})
\begin{equation}\label{result-lambda5}
	 \Delta  _{H_2}(z_{\lambda_5})= -\frac{16 k \left(z_{\lambda_5}\right) }{158+13 k \left(z_{\lambda_5}\right)}\Delta  _{B-L}\,.
\end{equation}
For practical reasons, it is convenient to define
\begin{equation}
\Delta_{H_2}^{\lambda_5}\equiv | \Delta  _{H_2}(z_{\lambda_5})| \ .
\end{equation}
Let us note that since $z_{\lambda_5}\gg1$,  the $k$-factor can be approximated by
\begin{equation}
\label{klimit}
k\left(z_{\lambda_5}\right){\simeq}  12\ \left(\frac{z_{\lambda_5}}{2\pi  }\right)^{3/2} \ e^{-{z_{\lambda_5}}} \,.
\end{equation}
Clearly, the 
$\lambda_5$ coupling must not be too large in order to avoid a too strong exponential suppression of the $H_2$ asymmetry.

As emphasized in Ref.~\cite{Servant:2013uwa}, for fermion quartic interactions,  the last $\lambda_5$ induced channels to decouple are the t-channel ones,\footnote{We thank G.~Servant and S.~Tulin for discussions on the importance of these channels.} simply because they are less Boltzmann suppressed than the other ones. 
 The value of $z_{\lambda_5}$ is   given by the condition that the 
$\Gamma_{\lambda_5}$ rate
is equal to the Hubble rate $H$.  One has for example that if $m_{H_1}=10$~TeV,  $z_{\lambda_5}=2(15)$ for $\lambda_5=3\cdot10^{-6}(5\cdot10^{-6})$. For  somewhat smaller values of $\lambda_5$, even if the $\lambda_5$ reactions doesn't enter in thermal equilibrium (as defined by Eq.~(\ref{lambda5-condition})), a numerical integration of the Boltzmann equations shows that still a number of scattering processes occur nevertheless, what can lead to a sizable asymmetry.

\subsection{ $\mathbf{T_{\lambda_5}>T>T_{fo}}$}

During this period, the $\Delta_{H_2}$ asymmetry stays constant unlike the total 
abundance $\Sigma_{H_2}$,  whose Boltzmann equation for this period reads
\begin{equation} \label{boltzmann-equation-zfo}
	\frac{d \Sigma_{H_2}}{dz} =-   \frac{\langle\sigma_{eff}  v\rangle s}{zH}    \left[  \Sigma_{H_2}^2 -\Delta_{H_2}^{\lambda_5\ 2} 	  -\Sigma_{H_2}^{eq\ 2}\right]\ ,
\end{equation}
where $\langle \sigma_{eff} v \rangle$ is the effective thermal cross section of the $H_2\bar H_2\leftrightarrow SM\, SM$ annihilations, given in the Appendix.  With  a constant $\Delta_{H_2}^{\lambda_5}$, as it is the case during this period, the solution of Eq.~(\ref{boltzmann-equation-zfo}) at freeze-out is to a good approximation given by
\begin{equation}\label{result-inter}
 \Sigma_{H_2}(z_{fo})\simeq \left[\Delta_{H_2}^{\lambda_5\ 2} +\Sigma_{H_2}^{eq\ 2}\left(z_{fo}\right) \right]^{1/2}\ ,
\end{equation}
which is nothing but the expression which makes the r.h.s. of Eq.~(\ref{boltzmann-equation-zfo}) to vanish.
Here, by $z_{fo}$ we mean  the usual freeze-out value given by the equation
\begin{equation}
z_{fo}\simeq \ln \frac{0.0038\cdot m_{Pl}\, 2 g_{H_2}\, m_{H_2}  \langle \sigma_{eff} v \rangle}{\sqrt{g_*z_{fo}}}     \,.
\end{equation}
If the annihilations are fast enough to leave at $T_{fo}$ a symmetric component smaller than the asymmetric one (which is typically satisfied for $\langle \sigma_{eff} v\rangle \gtrsim 1$~pb), the following relation holds, $\Sigma_{H_2}(z_{fo})\simeq  \Delta_{H_2}^{\lambda_5}  \gg  \Sigma_{H_2}^{eq}(z_{fo})$.  Given the sign of the baryon asymmetry, this means at $T_{fo}$, $\Sigma_{H_2}\sim -\Delta_{H_2}\sim -Y_{\bar H_2}\gg Y_{H_2},\Sigma_{H_2}^{eq}$.

\subsection{$\mathbf{T_{fo}>T>T_{EW}}$}

Nothing is expected to happen during this period. The $H_2$ total density left at $T_{fo}$ is left intact until $T_{EW}$, temperature at which  the total density and asymmetry are given by (for a dominant asymmetric component)  
\begin{align}\label{sigma-EW}
 \Sigma_{H_2}(z_{EW})\simeq \Sigma_{H_2}(z_{fo})\simeq \Delta_{H_2}^{\lambda_5}   \quad\text{and}\quad	 |\Delta_{H_2}(z_{EW})|=\Delta^{\lambda_5}_{H_2}	\, .
\end{align}

\subsection{ $\mathbf{T<T_{EW}}$}

Next, once the temperature drops below $T_{EW}$, two new effects   enter into play: generation of mass splittings between the $H^0$, $A^0$ and $\eta^+$ components and possibly fast inert particle-antiparticle oscillations $\eta^0\leftrightarrow \eta^{0*}$.
The effect of the mass splittings generated by the SM scalar vev, Eq.~(\ref{mass-splitting}), is of moderate importance. Assuming, as said above, that the $H^0$ component is the lightest one (i.e.~$\lambda_5<0$), they imply that the other components will ultimately decay to $H^0$.
But these decays conserve the number of inert scalar particles. They just convert the $H_2$ asymmetry created before EWSB (with mass $m_{H_2}$) into a DM relic density of selfconjugated  DM particles $H^0$ (with mass $m_{H^0}=m_{DM}$, different from $m_{H_2}$ unless $\lambda_{H^0}$ vanishes). 
More important is the potential effect of the much faster inert particle-antiparticle oscillations $\eta^0\leftrightarrow \eta^{0*}$  caused by the $\lambda_5$ interactions. The rate of an oscillating particle is simply given by the value of the associated mass splitting \cite{Arina:2011cu,Tulin:2012re,Cirelli:2011ac},  i.e.~$\Gamma_{osc}=\delta m=m_{A^0}-m_{H^0}$. For $T<T_{EW}$, this rate is very fast compared to the Hubble rate
 \begin{equation}\label{oscillation-rate}
 \frac{\Gamma_{osc}}{H\left(T\right)}  \simeq 2\cdot 10^{15} \cdot|\lambda_5|\cdot\left(\frac{100\,\hbox{GeV}}{T}\right)^2\cdot\left(\frac{\hbox{TeV}}{m_{H^0}}\right)  \ .
 \end{equation}
The effect of these oscillations depends  obviously  on whether they do occur, which Eq.~(\ref{oscillation-rate}) doesn't necessarily imply, and on whether symmetric annihilations do occur after EWSB.
Actually, even if the freeze-out occurs before EWSB, this does not imply that symmetric annihilations could not restart again after EWSB, due to oscillations  \cite{Cirelli:2011ac,Tulin:2012re}.
This could easily be the case because, even if one starts with a pure asymmetry, the oscillations will quickly give a number density of each population much larger than their thermal equilibrium values,  roughly $n_{\eta^0}\sim n_{\eta^{0*}}\sim| \Delta n_{\eta^0}|/2\gg n_{\eta^0}^{eq}$,
so that $|\Delta n_{\eta^0}| \langle\sigma v\rangle> H$ can hold even if $n_{\eta^0}^{eq} \langle\sigma v\rangle< H$. 
If these annihilations occur, they will anyway reduce the DM abundance, as no inverse processes will occur in this case. Let us consider both possible cases separately.

\subsubsection{$\mathbf{T<T_{EW}}$:  No symmetric annihilations after EWSB } 

If no symmetric annihilations arise after EWSB, oscillations have simply no effect. They quickly reconvert a pure $\eta^0$ population, or a pure $ \eta^{0*}$ population, into an oscillating mixed $\eta^0$-$\eta^{0*}$ population, but they do not change the number of inert states \cite{Cirelli:2011ac}. 
In this case, the number of $H^0$ particles left today will be simply equal to the number of inert scalar particles stored in the $H_2$ asymmetry before EWSB, i.e.~$Y_{DM}^{today}\simeq \Delta_{H_2}^{\lambda_5}  $,
that means the $H^0$ density is equal to the asymmetry left after $\lambda_5$ interaction's decoupling.
From  Eqs.~(\ref{result-lambda5})  and (\ref{sigma-EW}),  this gives
 \begin{equation}\label{H2-fct-B-L}
Y_{DM}^{today}=\frac{16 k \left(z_{\lambda_5}\right) }{158+13 k_{H_2}\left(z_{\lambda_5}\right)}\Delta_{B-L} \ ,
 \end{equation} 
with $k(z_{\lambda_5})$ given by Eq.~(\ref{klimit}).
Only the relation between the value of $Y_B$ today and  $\Delta _{B-L}$ changes after EWSB, as a result of the fact that below $T_{EW}$ the conservation of electric charge holds rather than conservation of  $Y$ and  $T_3$. We get 
\begin{equation}\label{B-below-EW-high-T}
    Y_B^{today}\simeq\frac{12}{37}\Delta_{B-L} \ .
\end{equation}
As a result, the DM density  reads
\begin{equation}\label{YDM-no-sym}
	Y_{DM}^{today}= \frac{148\cdot k\left(z_{\lambda_5}\right) }{474+39\cdot k\left(z_{\lambda_5}\right)}\cdot Y_B^{today}\ ,
\end{equation}
and the actual DM to baryon density ratio is given by
\begin{align}\label{eq:DM-abundance-high-T}
	  \frac{\Omega_{DM} }{\Omega_{B} } = \frac{Y^{today}_{DM}  }{Y^{today}_{B} }\cdot\left(\frac{m_{H^0}}{ \text{1 GeV}} \right)= \frac{148\cdot k\left(z_{\lambda_5}\right) }{474+39\cdot k\left(z_{\lambda_5}\right)} \cdot  \left(\frac{m_{H^0}}{\text{1 GeV}}  \right)\ .
\end{align}
This is the final result if no symmetric annihilations occur after EWSB.  
 However,  it must be stressed that it is not mandatory to avoid symmetric annihilations after EWSB. On the contrary, if the $\lambda_5$ interaction above does not provide enough suppression, these scattering processes could easily provide it, without the need of any special tuning.  This is what we will now quantify.

\subsubsection{$\mathbf{T<T_{EW}}$: Symmetric annihilations after EWSB}  

Possible effects of dark matter oscillations have been studied in Refs.~\cite{Cirelli:2011ac,Tulin:2012re}. As said above, since  oscillations reprocess the asymmetry into both particle and antiparticle densities,    their main effect is to allow the symmetric annihilations to start again.
Even if, as Eq.~(\ref{oscillation-rate}) shows, the oscillation rate is much faster than the Hubble rate, this doesn't necessarily mean that oscillations (and thus eventually annihilations) do occur. 
As shown in these references, if dark matter undergoes fast annihilations or elastic scatterings, these processes can break the coherence of the $\eta^0$-$\eta^{0*}$ states, preventing them from oscillating. The interplay of the oscillations with the other processes is actually more complicated in our doublet scenario than in the singlet setups considered in Refs.~\cite{Cirelli:2011ac,Tulin:2012re}. On top of $\eta^0$-$\eta^{0*}$ annihilations and $\eta^0$ or $\eta^{0*}$ elastic scatterings, there are charged $\eta^\pm$ states,  which at $T_{EW}$ are responsible for half of the asymmetry and do not oscillate. Fast inelastic scatterings can  change neutral states into these charged states and vice et versa. Moreover, as said above, all states ultimately become real $H^0$ states, which obviously do not oscillate.

 Let us first consider what happens to the neutral states, 
as if there were no charged states. 
In this case, one has two important processes. On the one hand, there are $\eta^0 \eta^{0*}\rightarrow SM SM$ annihilations processes which are dominated by their  $\lambda_{3,4}$ interaction contribution. On the other hand there are $\eta^{0(*)} SM \rightarrow \eta^{0(*)} SM$ elastic scatterings which are dominated by their t-channel $Z$ exchange contribution.
The later dominates over the $\lambda_{3,4}$ elastic scattering contribution.  For $T\gtrsim m_h$, this stems from the fact that
it involves a t-channel mediator whose mass is much smaller than the inert scalar mass, $m_Z\ll m_{H_2}$. 
It scales as $\Gamma^{gauge}_{scat}\simeq G_F^2T^5$ as compared to the quartic coupling elastic contribution which scales as $\Gamma_{scat}^{quartic}=n_{H_1}^{eq}\langle \sigma v \rangle \simeq \lambda_{3,4}^2 T^3/m_{H_2}^2$.
 For $T<m_h$, the quartic contribution is also subleading because it is Boltzmann suppressed, unlike the gauge one. 
As a result, the  gauge elastic contribution is the last to decouple. 
The $Z$ exchange process is relevant for preventing $\eta^0$-$\eta^{0*}$ states from oscillating because the gauge interaction is odd under $\eta^0$-$\eta^{0*}$ exchange \cite{Tulin:2012re}. 
Thus the relevant question is, down to which temperature will these processes effectively prevent the oscillations to start~? At first sight, we could think  that oscillations will start only once the scattering rate $\Gamma_{scat}$ goes below the oscillation rate $\Gamma_{osc}=\delta m$ (if at this time both rates are still larger than the Hubble rate). This turns out to occur at a rather low temperature, $T_{osc}\sim$~few~GeV scale. In this case one would be back to the ``no symmetric annihilation'' case above,
 because  oscillations have practically no more effect at this temperature,  where the  annihilation rate is already largely suppressed. 
However, an integration of the Boltzmann equations shows that oscillations rather start when  $(\delta m)^2/H=\Gamma_{scat}$,  see Ref.~\cite{Cirelli:2011ac}. In our scenario, as we also have checked from a numerical integration of the relevant Boltzmann equations, this turns out to happen at a temperature above $T_{EW}$.   Thus we conclude that oscillations start as soon as EWSB occurs. As a result, annihilations can restart from this temperature and to determine how much of them will annihilate, one can just take the Boltzmann equations with the oscillation and annihilation terms,
\begin{align}\label{Boltzmann-equation-1-1}
	\frac{d\Sigma_{\eta^0}}{dz}&=-\frac{\langle\sigma_{0 } v\rangle\ s}{2zH }\left[	  \Sigma^2_{\eta^0} -\Delta^2_{\eta^0} -\Xi^2_{\eta^0}   -\Sigma^{eq\ 2}_{\eta^0} 	\right]\ ,\\\label{Boltzmann-equation-1-2}
	\frac{d\Delta_{\eta^0}}{dz}&=2 i\frac{ \delta m}{zH } \ \Xi_{\eta^0} \ ,\\\label{Boltzmann-equation-1-3}
	\frac{d\Xi_{\eta^0}}{dz}&=2 i\frac{\delta m}{zH }\ \Delta_{\eta^0}-\frac{\langle\sigma_{ 0}  v\rangle s}{zH}\ \Xi_{\eta^0}\ \Sigma_0 \ ,
\end{align}
where for any $T\leq T_{EW}$ we define $z\equiv m_{H^0}/T$,  with $\langle\sigma_{ 0}  v\rangle$ the  thermally averaged $\eta^{0}\eta^{0*}\to SM$ annihilation cross section, 
and $\Xi_{\eta^0}$ a quantity that accounts for the coherence between the $\eta^0$ and $\eta^{0*}$ components (see \cite{Cirelli:2011ac} for further details). The resolution of these equations leads to a monotonically decreasing $\Sigma_{\eta^0}(z)$ function and to  oscillating functions $\Delta_{\eta^0}(z)\propto \cos[f(z)]$ and $\Xi_{\eta^0}(z)\propto \sin[f'(z)]$ whose amplitudes also decrease  monotonically.
For fast oscillations, and neglecting the $\Sigma^{eq}_{\eta^0}$ term in Eq.~(\ref{Boltzmann-equation-1-1}), the set of Boltzmann equations can be simplified and solved analytically, at an approximate level, as explained in the Appendix.
The solution it gives for $\Sigma_{\eta^0}$ is~\footnote{This result is approximately the same  than the one obtained in \cite{Cirelli:2011ac} for much smaller $\delta m$ values -- see Eqs~(25) and (33) therein -- but in which $x_{osc,ann}$ (which depends on $\delta m$) is now simply replaced by $z_{EW}$.  } 
\begin{align}\label{solution-sigma}
\Sigma_{\eta^0}(z\geq z_{EW})=\frac{\Sigma_{\eta^0}(z_{EW}) }{  1+\frac{1}{2} \frac{\langle\sigma_{ 0}  v\rangle  \, s(z)}{H(z)} \left(\frac{z}{z_{EW}}-1\right)\Sigma_{\eta^0}(z_{EW})} \ , 
\end{align}
with  $z_{EW}=m_{H^0}/T_{EW}$ and where  we fixed  the initial abundance and asymmetry to be equal to $\Sigma_{\eta^0}(z_{EW})= \Delta_{H_2}^{\lambda_5} /2$. The asymmetry $\Delta_{\eta^0}$ and $\Xi_{\eta^0}$ are, in turn, fast oscillatory functions which are equal to zero on average.

The result of Eq.~(\ref{solution-sigma}) can also be qualitatively understood in the following way. Once $T\leq T_{EW}$, the fast oscillations reprocess quasi instantaneously the  $\eta^{0}$ asymmetry  in oscillatory abundances for $\eta^0$ and $\eta^{0*}$.  On average, just after EWSB, we have therefore $n_{\eta^0}\simeq n_{\eta^{0*}}\simeq|\Delta n_{\eta^0}|_{T_{EW}}/2$.  Since 
$T_{fo}>T_{EW}$, 
when two conjugate particles annihilate to SM particles, the reduction of inert doublet state it implies will not be compensated by any
inverse processes. As a result, the Boltzmann equation for $\Sigma_{\eta^0}$ one gets along this way is  simply given by
\begin{equation}\label{beq-naive-approach}
	\frac{d \Sigma_{\eta^0}}{dz} =- \frac{\langle\sigma_{0 }  v\rangle s}{2zH}  \ \Sigma_{\eta^0}^2\ ,
\end{equation}
whose resolution leads to nothing else than Eq.~(\ref{solution-sigma}).\footnote{The reason why the two results coincide is in fact more subtle. 
 Since   the   $\eta^{0(*)}$  oscillatory behavior is given by \[Y_{\eta^{0(*)}}=\frac{1}{2}f(z)\left(1\pm\cos g(z)\right)\ ,\] 
 the Boltzmann equation in this naive approach should read
 \[\frac{d \Sigma_{\eta^0}}{dz} =- 2 \frac{\langle\sigma_{ 0}  v\rangle s}{zH}  \ Y_{\eta^0}Y_{\eta^{0*}}=- \frac{\langle\sigma_{0 }  v\rangle s}{2zH}  \ \Sigma^2_{\eta^0}\ \sin^2 g(z) \ .\]
Averaging this expression, we find Eq.~(\ref{beq-naive-approach}) up to an extra factor $1/2$. An extra factor 2 must nevertheless be added 
to take into account the contribution of the  coherence $\Xi_{\eta^0}$ part, giving back Eq.~(\ref{solution-sigma}). }

The next step is to include the contribution of the charged states. Since these states do not oscillate, one could naively expect that the charged asymmetry is essentially left intact until the charged states decay to $H^0$ states.
This doesn't work this way. To see that precisely, one should in principle solve the corresponding set of   six coupled Boltzmann equations, for $\Sigma_{\eta_0}$, $\Delta_{\eta_0}$, $\Xi_{\eta_0}$, $\Pi_{\eta_0}$, $\Sigma_{\eta_+}$, $\Delta_{\eta_+}$,  where $\Pi_{\eta_0}$ and  $\Xi_{\eta_0}$ are the real and imaginary parts of the quantity that accounts for the coherence effects. Nevertheless in practice we don't need to go that far.
It turns out that, if  just before EWSB there are essentially only $\eta^-$ and $\eta^{0*}$ states as considered here,
as soon as oscillations start they put the neutral state asymmetry to zero (on average), and processes which can transfer a charged asymmetry into a neutral one will very quickly put the charged asymmetry to zero too.
This will be done in particular by $\eta^{0(*)}\, SM\leftrightarrow\eta^{\pm}\, SM$ inelastic scatterings  and $\eta^\pm\leftrightarrow \eta^{0(*)} SM SM$ decays. The decrease of the charged component asymmetry due to these processes is exponential ($\Delta_{\eta^+}\propto e^{-z}$, as can be seen from the corresponding term in the Boltzmann equation, $sHz\, d\Delta_{\eta^+}/dz\propto -\Delta_{\eta^+}\gamma_{\eta^+\leftrightarrow \eta^0}+...$).
This ``re-equilibration''   of the asymmetries by these processes, which follows  their ``desilagnement'' by  the oscillations when these  latter 
start, occurs much faster than the process of suppression of $\Sigma_{\eta_0}$ in Eq.~(\ref{Boltzmann-equation-function}).
As a result, in the same way as for the neutral states, 
one can adopt the simple assumption that   as soon as oscillations start, the particle and antiparticle densities for charged states are equilibrated, $Y_{\eta^0}=Y_{\eta^{0*}}=Y_{\eta^+}=Y_{\eta^-}$.   
At this point, the  annihilation processes such as $\eta^+\eta^{-}\to SM SM$, $\eta^+\eta^{0*}\to SM SM$ and $\eta^-\eta^{0}\to SM SM$  can start again, in the same way as the $\eta^0\eta^{0*}\to SM SM$ ones. The whole effect 
can be approximatively accounted by the simple Boltzmann equation
\begin{equation}\label{Beq-naive-approach-final}
	\frac{d \Sigma_{H_2}}{dz} =- \frac{\langle\sigma_{eff}  v\rangle s}{zH}  \ \Sigma^2_{H_2}\,.
\end{equation}
Similarly to what has been obtained in Eq.~(\ref{solution-sigma}),  the resolution of Eq.~(\ref{Beq-naive-approach-final}), integrated from $T_{EW}$ until now and using the initial condition in (\ref{sigma-EW}), leads to 
\begin{equation}\label{eq:SigmaH2-kappa-suppression}
\Sigma_{H_2}(z\geq z_{EW})=\frac{ \Delta_{H_2}^{\lambda_5} }{  1+ \frac{\langle\sigma_{eff}  v\rangle  \, s(z)}{H(z)} \left(\frac{z}{z_{EW}}-1\right) \Delta_{H_2}^{\lambda_5} }
\ .
\end{equation}
This equation holds for the case where the total number density just before EWSB is given by the asymmetry. If there is also a non-negligible part which is left from the symmetric freeze-out, one must simply replace the asymmetry  at $z_{EW}$,  $\Delta_{H_2}^{\lambda_5} $, by the total number density at the same temperature, $\Sigma_{H_2}(z_{EW})$, since this is the number which determines the number of symmetric annihilations which will occur after EWSB,\footnote{If there is no asymmetry and if the freeze-out has occurred prior to EWSB, one recognizes in Eq.~(\ref{eq:SigmaH2-kappa-suppression-total}) the usual asymptotic freeze-out behavior, i.e.~the freeze-out is not instantaneous, but reaches asymptotically its final value as given in this equation. 
In practice, as well known,
the effect is negligible in this case, i.e.~the denominator is equal to unity to a good approximation. Here, instead, the denominator at $z_{fo}$ can be much larger due to the asymmetry. 
}
\begin{equation}\label{eq:SigmaH2-kappa-suppression-total}
\Sigma_{H_2}(z\geq z_{EW})=\frac{ \Sigma_{H_2}(z_{EW})}{  1+ \frac{\langle\sigma_{eff}  v\rangle  \, s(z)}{H(z)} \left(\frac{z}{z_{EW}}-1\right)\Sigma_{H_2}(z_{EW}) }
\ .
\end{equation}
with $\Sigma_{H_2}(z_{EW})\simeq \Sigma_{H_2}(z_{fo})$ as given in Eq.~(\ref{result-inter}).

Note also that the $\eta^0\, SM\leftrightarrow\eta^+\, SM$ (and conjugated) processes above not only equilibrate the neutral and charged asymmetries, but also can break the coherence of the $\eta^0$-$\eta^{0*}$ by transforming a coherent neutral state into a charged state which does not oscillate.  However, in the same way as for the $Z$ exchange channel above, its rate goes under $(\delta m)^2/H$ before EWSB occurs, so that they do not prevent oscillations to start at $T=T_{EW}$.

Finally, because of the mass splittings between $H^0$ and the other components of the inert doublet, this total density is progressively transferred into a $H^0$ density  through the decays of  the heavier components. 
Note that, as we will see, in the numerical section below, Eq.~(\ref{eq:SigmaH2-kappa-suppression-total}) reaches its asymptotic value to a good approximation before $T$ drops below the value of the mass splitting $m_{A^0}-m_{H^0}$.
As a result, this splitting can be neglected as it was done to get Eq.~(\ref{eq:SigmaH2-kappa-suppression-total}). 

 \begin{figure}
\centering
    \includegraphics[scale=.65]{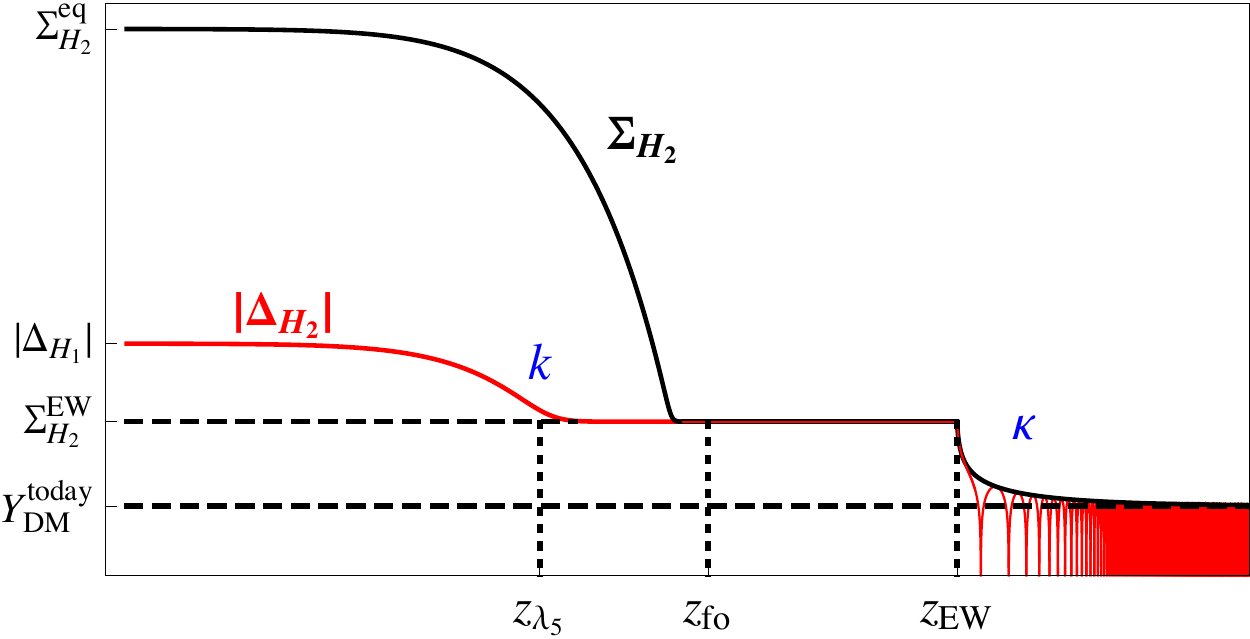}   
  \caption{Sketch of the scenario  considered in section II and III. We represent, as a function of  $z=m_{H^0}/T$,  the $H_2$ asymmetry $|\Delta_{H_2}|$ in red and the total DM density $\Sigma_{H_2}$ in black. We neglect in this sketch the mass splittings between the different components of $H_2$. 
First step:  The initial asymmetry $\Delta_{H_1}$ and $\Delta_{H_2}$ are fixed by the $B-L$ asymmetry. Second step: this asymmetry gets suppressed until the $\lambda_5$ interactions decouple at $z_{\lambda_5}$ --- the suppression is characterized by the $k$-factor. Third 
step: at $z_{\lambda_5}$ the total annihilation cross section is still in thermal equilibrium and $\Sigma^{eq}_{H_2}$ follows  
the equilibrium density $\Sigma^{eq}_{H_2}$ until it reaches $|\Delta_{H_2}|$. Fourth step: at EWSB, oscillations start and reequilibrate the particle-antiparticle populations.  At this point, annihilations can start again and if they do they deplete the density -- the suppression is characterized by the $\kappa$-factor. }
\label{sketch}
\end{figure}

\subsection{Final inert scalar relic density}

We summarize in Fig.~\ref{sketch} the evolution of the asymmetry $|\Delta_{H_2}|$ and total density $\Sigma_{H_2}$. We remind the main steps:
\begin{enumerate}
\item   $T\gtrsim m_{H_2}$. The $H_2$ asymmetry,  proportional to the $B-L$ asymmetry, is generated through the $\lambda_5$ interactions:
   \begin{equation}
	\Delta _{H_2}(z\lesssim1)= -\frac{4}{23}\Delta _{B-L} \,. \tag{\ref{eq:eq-cond-fast-mH2}} 
\end{equation}
\item $ m_{H_2}\gtrsim T \geq T_{\lambda_5}$. The asymmetry undergoes a Boltzmann suppression until the $\lambda_5$ interaction decouples,
 \begin{equation}  \tag{\ref{result-lambda5}} 
 	\Delta^{\lambda_5}_{H_2}\equiv |\Delta  _{H_2}(z_{\lambda_5})|= \frac{16 k \left(z_{\lambda_5}\right) }{158+13 k \left(z_{\lambda_5}\right)}\Delta  _{B-L}\ .
\end{equation}
\item  $ T_{\lambda_5}> T > T_{EW}$. The symmetric component of  $\Sigma_{H_2}$ follows an exponential suppression, until 
$\Sigma_{H_2}$ reaches  at $z=z_{fo}$ the value
\begin{equation}  \tag{\ref{result-inter}} 
\Sigma_{H_2}(z_{fo})\simeq \left[\Delta_{H_2}^{\lambda_5\ 2} +\Sigma_{H_2}^{eq\ 2}\left(z_{fo}\right) \right]^{1/2} \ .
\end{equation}
From $z_{fo}$ to $z_{EW}$, the annihilations are momentarily  frozen, so that $\Sigma_{H_2}(z_{EW})\simeq\Sigma_{H_2}(z_{fo})$.
During this period, if the contribution from usual freeze-out is negligible, there are only $\bar H_2$ particles in the plasma and $\Sigma_{H_2}(z_{EW})\simeq \Delta^{\lambda_5}_{H_2} $.
\item  $T < T_{EW}$. The fast $\eta^0 \leftrightarrow \eta^{0*}$ oscillations start. They quickly reprocess the $\bar H_2$ density in equal abundances (on average) for $\eta^0$, $\eta^{0*}$, $\eta^+$ and $\eta^-$. The annihilations can therefore start again and, if they do, they  deplete the set of densities, whose sum reads asymptotically   
\begin{equation}    
\Sigma_{H_2}(z\gg z_{EW})=\frac{\Sigma_{H_2}(z_{EW})}{  1+ \frac{\langle\sigma_{eff}  v\rangle  \, s(z)}{H(z)} \frac{z}{z_{EW}} \Sigma_{H_2}(z_{EW}) } \ .
\end{equation}
This total density is progressively transferred into a $H^0$ density through the decay of  the heavier components.
\end{enumerate}
The final  DM abundance is  therefore given by 
\begin{align}\label{eq:DM-abundance-high-T-sym}
	Y_{DM}^{today}=\Sigma_{H_2}(z\gg z_{EW})=\frac{\Sigma_{H_2}(z_{EW})}{1+ \kappa\cdot \Sigma_{H_2}(z_{EW})}  \ ,
\end{align}
where we define 
\begin{equation}\label{kappa-factor}
\kappa\equiv \frac{\langle\sigma_{eff}  v\rangle  s(z) }{H(z)} \frac{z}{z_{EW}}  \simeq 1.3\cdot 10^{13}\cdot \left(\frac{\langle\sigma_{eff}  v\rangle  }{ 1\text{ pb}} \right)  \ .
\end{equation}
Since ultimately no asymmetry survives, the relation between the baryon and the $B-L$ asymmetry is still  given by Eq.~(\ref{B-below-EW-high-T}), and  the final DM to baryon density  ratio is given by
\begin{align}\label{eq:DM-final}
	  \frac{\Omega_{DM} }{\Omega_{B} } &=\frac{\Sigma_{H_2}(z_{EW}) }{1+ \kappa\cdot\Sigma_{H_2}(z_{EW})} \cdot\frac{1}{Y_B^{today}}\cdot\left(\frac{m_{H^0}}{\text{1 GeV}}\right) \ ,
\end{align}
or equivalently, if the asymmetric component dominates, using Eqs.~(\ref{result-lambda5}) and (\ref{sigma-EW}),
\begin{align}\label{eq:DM-final-bis}
	  \frac{\Omega_{DM} }{\Omega_{B} } &=   \frac{148 \, k\left(z_{\lambda_5}\right) }{474+\left(39+148\, \kappa\,  Y_B^{today}\right)  k\left(z_{\lambda_5}\right)}  \cdot \left(\frac{m_{H^0}}{\text{1 GeV}}\right) \ .
\end{align}

A number of comments can be done regarding these results:
\begin{itemize}
\item Eqs.~(\ref{eq:DM-abundance-high-T-sym}) and (\ref{eq:DM-final}) show that beside the $\lambda_5$ interaction induced ``$k$-factor" suppression in $\Delta_{H_2}$, see  Eq.~(\ref{result-lambda5}), oscillations drive a $1+ \kappa\cdot\Sigma_{H_2}(z_{EW})$ factor suppression. This ``$\kappa$-factor'' suppression can be sizable as soon as $\kappa\cdot\Sigma_{H_2}(z_{EW})\gtrsim 1$.
\item As  Eq.~(\ref{eq:SigmaH2-kappa-suppression-total}) shows, this suppression is neither instantaneous nor exponential. It goes as the inverse of $z/z_{EW}-1$ until it reaches an asymptotic value. In this sense, imposing that the cross section satisfies the unitarity bound, it is naturally limited but it still  can be  responsible for  the $\sim(10\,\hbox{GeV}/m_{DM})$  suppression needed, see below.
\item  The appearance of the $\kappa\cdot \Sigma_{H_2}(z_{EW})$ factor is not surprising. The condition $\kappa\cdot\Sigma_{H_2}(z_{EW})<1$ is nothing but the condition $(n_{H_2}+n_{\bar{H}_2}) \langle\sigma_{eff}  v\rangle < H$ at $T=T_{EW}$.  
\item  Interestingly, for large values of $\kappa\cdot \Sigma_{H_2}(z_{EW})$, the $Y^{today}_{DM}$ relic density obtained doesn't depend anymore on the asymmetry left at $T_{EW}$, even if this asymmetry is the source of the final DM abundance. In this case, we simply get 
$$Y_{DM}^{today}=\frac{1}{\kappa}\,,$$
and the ratio reads
\begin{equation}
\label{eq:ratiofinallargekappa}
\frac{\Omega_{DM} }{\Omega_{B} } \simeq 0.15\cdot z_{EW}\cdot\left(\frac{ 1\text{ pb} }{ \langle\sigma_{eff}  v\rangle} \right) \ .
\end{equation}

This means, as we could have anticipated, that for large cross section the asymmetry left is independent of the initial asymmetry, provided this initial asymmetry is large enough. 
In other terms,  if the $\kappa$-factor suppression is small, both baryon and DM asymmetries are directly connected. If instead it is large, they are not related anymore in a so direct way, since in this case the final relic density depends only on the annihilation cross section.\footnote{But still, even in this case, they remain similar as the $\kappa$ factor is bounded from above by unitarity considerations on the total cross section.} 
Note interestingly that Eq.~(\ref{eq:ratiofinallargekappa}) is nothing but the result of the standard freeze-out scenario, but with the important difference that  in the standard case, $z_{EW}$ in Eq.~(\ref{eq:ratiofinallargekappa}) must be replaced by $z_{fo}$.

\end{itemize}

\section{\bf Failure of the asymmetric IDM scenario}   

\begin{figure}
\centering
    \includegraphics[scale=.63]{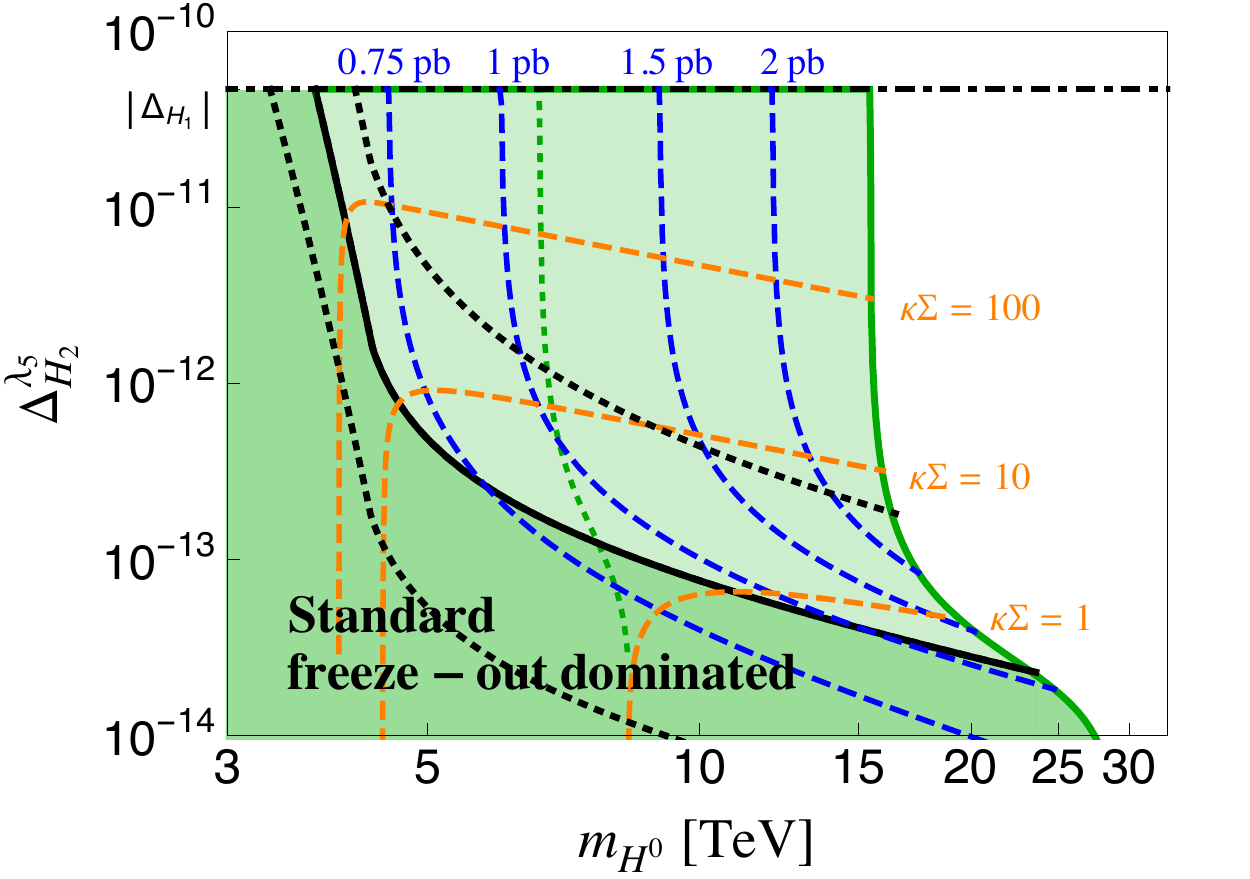}   
  \caption{Values of $\Delta_{H_2}^{\lambda_5}$ which give the observed relic density as a function of the input parameters $m_{H^0}$ for various values of
$\langle \sigma_{eff} v \rangle$ (dashed blue lines). The corresponding values of $\kappa\cdot\Sigma_{H_2}(z_{EW})$ are given by the dashed orange lines. The upper horizontal line gives the value of $\Delta_{H_2}^{\lambda_5}$ which is obtained from equilibration with the $H_1$ and $B-L$ asymmetries. The r.h.s.~solid (dashed) green line gives the maximum value of the input parameters imposing that $\lambda_{3,4}$ couplings are smaller than $4\pi$ ($\sqrt{4\pi}$).
Below $m_{H^0}\sim 4.7$~TeV, the freeze-out occurs after EWSB, which relatively quickly causes a huge suppression. The black lines from top to bottom give the value of the parameters  for which 90\%, 50\%, 10\% of the relic density is of asymmetric origin, respectively.}
\label{sigmaH2-vs-mass}
\end{figure}

The final result of Eq.~(\ref{eq:DM-final})  depends on three parameters:
$m_{H^0}$, $\Sigma_{H_2}(z_{EW})$ and the total cross section $\langle \sigma_{eff} v \rangle$ via $\kappa$ in Eq.~(\ref{kappa-factor}).  
This means that for given values of the input parameters  $m_{H^0}$ and $\langle \sigma_{eff} v \rangle$, there is
only one value of  $\Sigma_{H_2}(z_{EW})$ which gives the observed value of $\Omega_{DM}/\Omega_B$, as given by the PLANCK best fits, $\Omega_{DM}h^2=0.120$ and $\Omega_{B}h^2=0.022$ \cite{Planck:2015xua}. 
Since $\Sigma_{H_2}(z_{EW})$ depends only on  these two input parameters and on   $\Delta_{H_2}^{\lambda_5}$, this means also that there is only one value of $\Delta_{H_2}^{\lambda_5}$ which gives the correct relic density for fixed values of the two input parameters. We show in Fig.~\ref{sigmaH2-vs-mass} this value of  $\Delta_{H_2}^{\lambda_5}$ as a function of $m_{H^0}$ for different values of the cross section. By comparing this value of $\Delta_{H_2}^{\lambda_5}$ to the value this asymmetry would have if there were no ``$k$-factor'' suppression -- given by the $\Delta_{H_1}$ upper horizontal line -- one can read off what is the value of this $\lambda_5$ induced ``$k$-factor" suppression, Eq.~(\ref{result-lambda5}) as compared to Eq.~(\ref{eq:eq-cond-fast-mH2}). 

As said above, to dominate the final relic density, the asymmetry cannot be suppressed by more than a factor $m_{DM}/10$~TeV.
Figure~\ref{sigmaH2-vs-mass} also shows the corresponding values of the $\kappa\cdot \Sigma_{H_2}(z_{EW})$ factor which lead to the other suppression, i.e.~the $1/(1+\kappa\cdot \Sigma_{H_2}(z_{EW}))$ factor in Eq.~(\ref{eq:DM-final}). 
It also shows for which values of the various parameters the asymmetry produced before the EW transition is responsible for 50\% of the final DM relic density (black line). Above (below) this line the relic density is dominantly of asymmetric (symmetric) origin.
Similarly, the dotted upper (lower) black line gives the values of the parameters above (below) which the asymmetry is responsible for more (less) than 
90\% (10\%) of the final relic density. For masses which give a freeze-out below $T_{EW}$,
the $\kappa\cdot \Sigma_{H_2}(z_{EW})$ factor becomes exponentially large because in this case $ \Sigma_{H_2}(z_{EW})$ is still exponentially larger than its value at freeze-out. Thus, the proportion of $\Sigma_{H_2}(z_{EW})$ which is due to $\Delta_{H_2}^{\lambda_5}$ is therefore exponentially suppressed. This explains why the black lines quickly go up  for $m_{DM}$ below $4-5$~TeV. Note nevertheless that this suppression, even if exponential, is far from instantaneous. As a result we find that, still, the asymmetry can dominate the relic density for a mass equal to  3.7~TeV  which is substantially lower than the 4.7~TeV value which gives $T_{fo}=T_{EW}$.\footnote{To get this 3.7~TeV value we simply applied Eq.~(\ref{eq:SigmaH2-kappa-suppression}) neglecting the fact that in this case the $\Sigma_{H_2}^{eq\,2}$ inverse scattering term must be taken into account in the Boltzmann equations (as in Eq.~(\ref{Boltzmann-equation-1-1})). The incorporation of this term would slightly lower further this minimum value of $m_{DM}$.}
A comment which must be made at this point concerns the fact that 
we have considered the electroweak phase transition as if it was an instantaneous process, i.e.~as
a step function at the temperature $T_{EW}\sim 165$~GeV -- from Ref.~\cite{D'Onofrio:2014kta} (see also Ref.~\cite{Burnier:2005hp}) --
which as said above is the temperature where the vacuum expectation value of the SM scalar field becomes  sizable (i.e.~where the oscillations are about to start to reprocess the asymmetry). As the electroweak transition is a crossover, it is clearly an approximation which could be refined. A change of $T_{EW}$ by a given factor would shift all $m_{H^0}$ values in Fig.~\ref{sigmaH2-vs-mass} by about the same factor.

As expected from the discussion above, Fig.~\ref{sigmaH2-vs-mass} also shows that, for large value of $\kappa\cdot \Sigma_{H_2}(z_{EW})$, the observed relic density doesn't depend anymore on the value of  $\Delta_{H_2}^{\lambda_5}$, provided this later quantity is above a certain value.

Note that the r.h.s. green curve of Fig.~\ref{sigmaH2-vs-mass} is obtained by imposing that all quartic couplings are perturbative, $\lambda_{3,4}<4\pi$. 
This line shows that a dominant asymmetric component requires that $m_{DM}\lesssim 25$~TeV (whereas the same condition gives $m_{DM}\lesssim 30$~TeV for the standard freeze-out scenario and for a small value of the $\lambda_5$ coupling). Such a bound also implies an upper bound on the $\langle \sigma_{eff} v \rangle$ cross section of about 2.5 pb, that is to say a value about 4 times larger than the $\sim 0.7$~pb value one needs at these energies along the standard freeze-out scenario. Imposing instead that $\lambda_{3,4}<\sqrt{4\pi}$ one gets $m_{DM}\lesssim 8$~TeV and $\langle \sigma_{eff} v \rangle \lesssim 1.1$~pb (dashed green  line).

The minimum value of the $\lambda_3^2+\lambda_4^2$ coupling combination (which enters in $\langle \sigma_{eff} v \rangle$) that this scenario requires is $\sim 2$, corresponding to $m_{DM}\sim 4$~TeV and a cross section of $\sim0.5$~pb. This is smaller than the usual $\sim0.7$~pb   because  the associated asymmetry $\Delta^{\lambda_5}_{H_2}\sim|\Delta_{H_1}|$   also  participates to the depletion of the total density.  
No need to say that with such large values of these quartic couplings, Landau poles are to be typically expected far below the Planck scale.
Although the energy scale at which we get a Landau pole depends on the value of other couplings such as $\lambda_2$, if there is no cancellations between the contributions of various couplings in the beta functions, a value of $\lambda_{3,4}\sim 1.5$ gives a Landau pole at $\sim10^5$-$10^6$~GeV. This means that new physics is to be expected in this case below this value.
The scale of $B-L$ asymmetry production has not to be necessarily below this scale. All what matters for the value of $\Omega_{DM}/\Omega_B$ is the value of the $B-L$ asymmetry at $T\sim m_{H_2}$.

\begin{figure}
\centering    
\includegraphics[scale=.68]{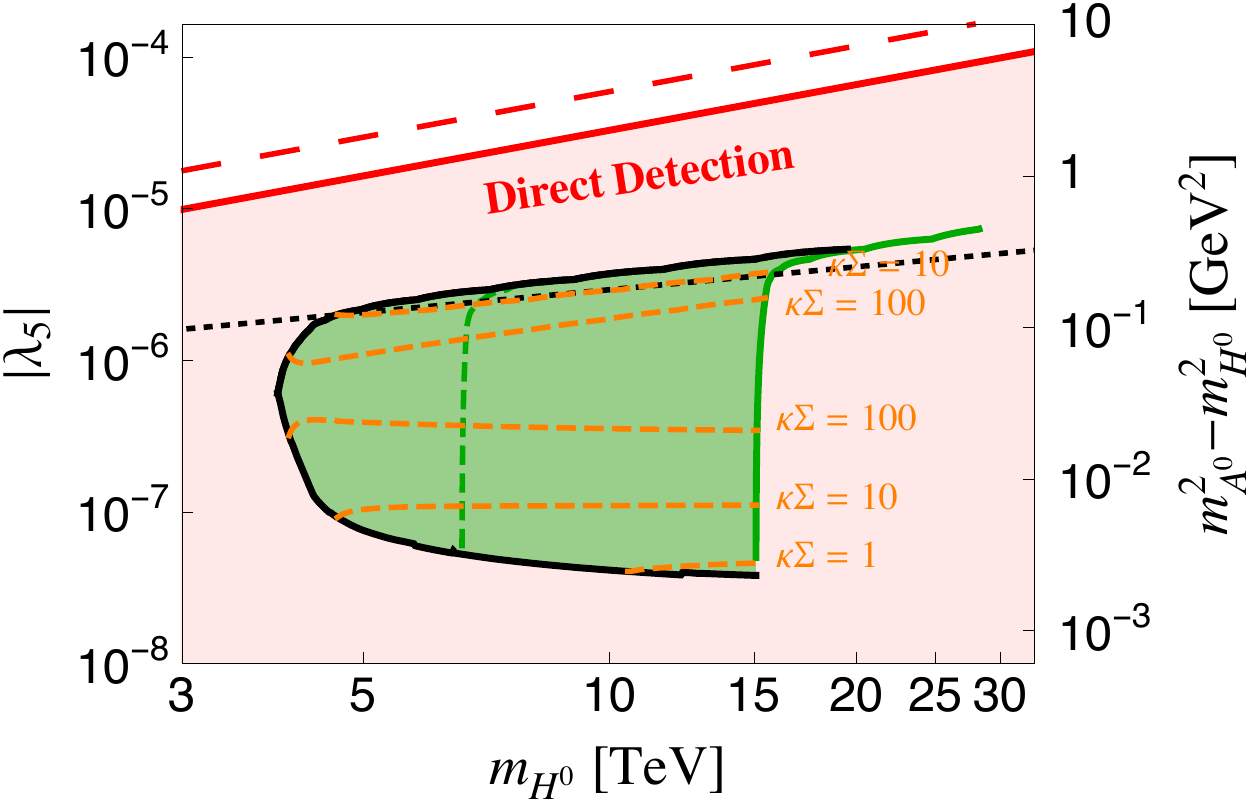}	  
\caption{Values of $\lambda_5$ which leads to the $\Delta_{H_2}^{\lambda_5}$ values needed in Fig.~\ref{sigmaH2-vs-mass}, as a function of the input parameters $m_{H^0}$. The red shaded area is excluded by the direct detection constraint of Eq.~(\ref{direct-detection}), taking $m_{A^0}-m_{H^0}=100$~keV. Taking instead $m_{A^0}-m_{H^0}=180$~keV gives the red dashed line. The dashed black line gives the value of $\lambda_5$ below which this interaction never gets in thermal equilibrium, as given by Eq.~(\ref{lambda5-condition}). The lowest allowed $\lambda_5$ value is obtained for a situation where there is neither a $k$ suppression, nor a $\kappa$ suppression.
In this case one generates directly the observed relic density from having only partial thermalization of the asymmetries.} 
\label{lambda5-vs-mH0} 
\end{figure} 

In Fig.~\ref{lambda5-vs-mH0}, as a function of the same two input parameters $m_{H^0}$ and $\langle \sigma_{eff} v \rangle$, we show the value of $\lambda_5$ which leads to the  $\Delta_{H_2}^{\lambda_5}$  value needed in Fig.~\ref{sigmaH2-vs-mass}. The corresponding value of the $m^2_{A_0}-m^2_{H_0}$ mass splitting is also given on Fig.~\ref{lambda5-vs-mH0}. 
This figure shows that the scenario leads to the observed relic density for  $\lambda_5\in[5\cdot 10^{-8},8\cdot10^{-6}]$, which corresponds to a mass splitting equal to approximately  $m_{A_0}-m_{H_0}\in[0.1,15]$~keV.
Larger values of $\lambda_5$ quickly lead to  a  $\lambda_5$ decoupling temperature much smaller than $m_{H_2}$, thus to largely Boltzmann suppressed remaining asymmetries. 
Smaller values of $\lambda_5$  rather  quickly lead to no thermalization of the $H_2$ and $H_1$ asymmetries, i.e.~to no creation of a $H_2$ asymmetry.  In most of the relic density allowed parameter space, both the ``$k$" and ``$\kappa$"  suppressions are  active, although it is possible to have only one of the effect to account for all the necessary suppression.
 As said above, an important constraint that one must satisfy is the direct detection constraint of Eq.~(\ref{direct-detection}). 
The value of the mass splitting just quoted are below the $\sim 100$~keV direct detection lower bound of Eq.~(\ref{direct-detection}).

Thus, unless direct detection would allow a mass splitting as low as  the value  $m_{A^0}-m_{H^0}\sim 15\,\hbox{keV}$, which seems very unlikely, this very minimal asymmetric scenario is in fact excluded~! 
This can also be clearly seen from Fig.~(\ref{lambda5-vs-mH0}) where the region allowed by direct detection  taking in Eq.~(\ref{direct-detection}) a mass splitting $\delta m_{min}=100$~keV has no overlap with the region which gives the observed relic density.
Or, in other words, imposing that the mass splitting is above $100\, \hbox{keV}$, the $\lambda_5$ interaction turns out to decouple only at $z_{\lambda_5}\gtrsim 50$ leading to a  tiny  $\Omega_{DM}$ relic density.

\section{Reprocessing the inert doublet asymmetry into a lighter particle DM relic density}
\label{sec:H2-decay}

 \begin{figure}
\centering
    \includegraphics[scale=.63]{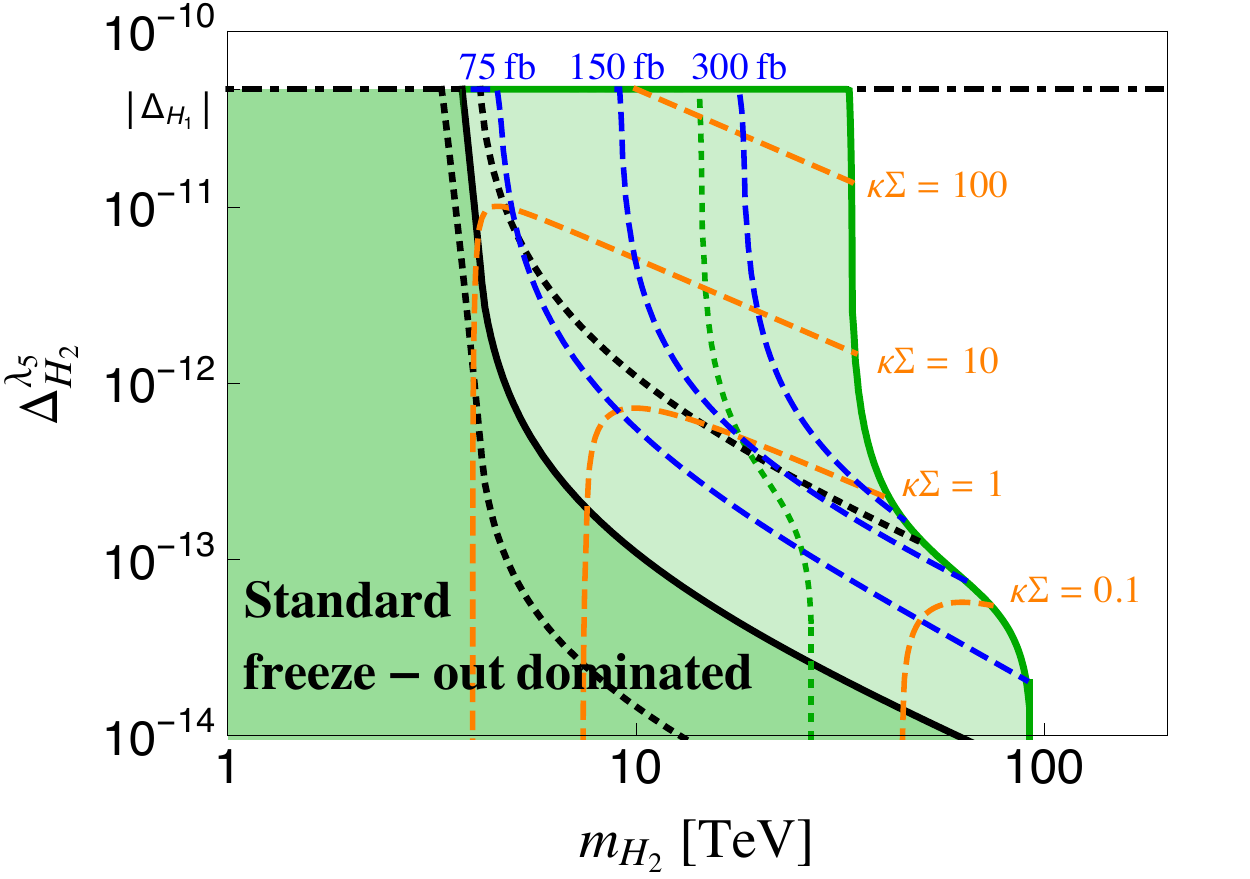}   
  \caption{
   Same as Fig.~\ref{sigmaH2-vs-mass}, but allowing the inert scalar to decay into a lighter real scalar singlet $S$ with mass $m_S=m_{H^0}/10$.}
\label{sigmaH2-vs-mass-bis}
\end{figure}

\begin{figure}
\centering
    \includegraphics[scale=.68]{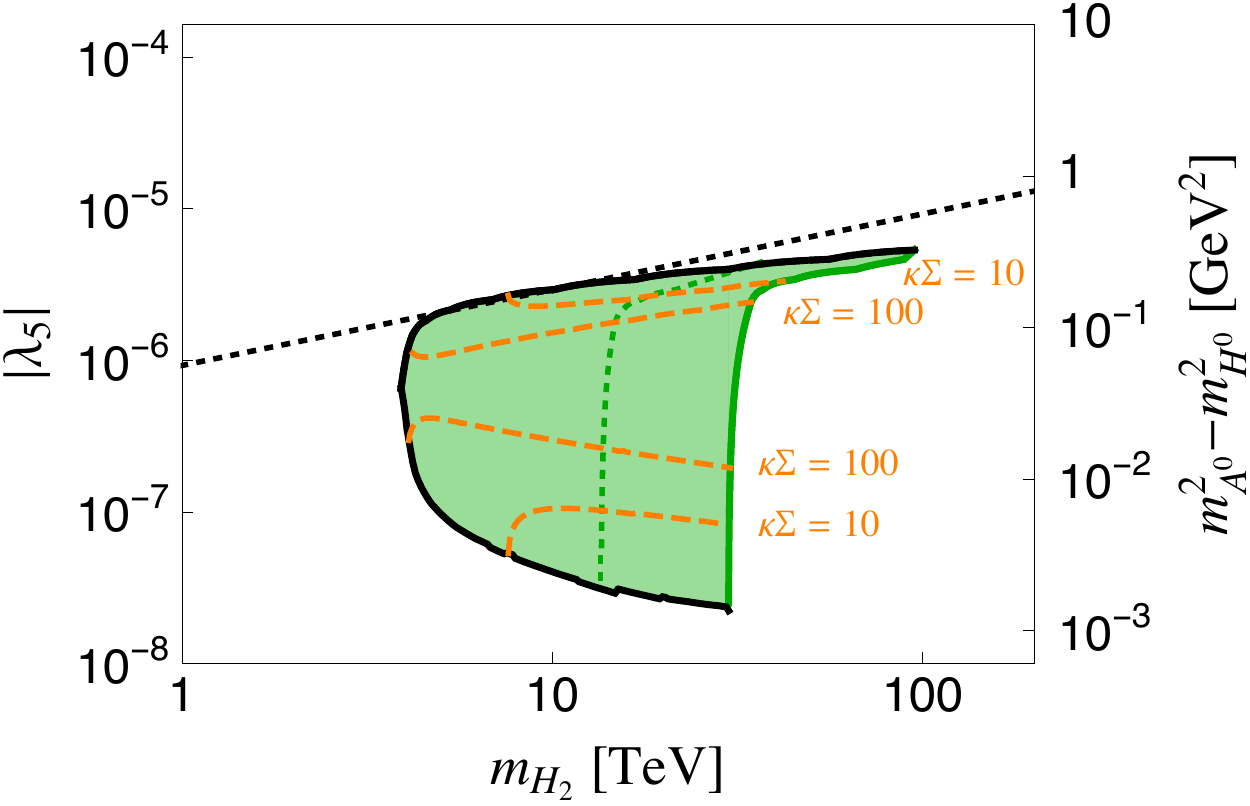}   
  \caption{
  Same as Fig.~\ref{lambda5-vs-mH0}, but allowing the inert scalar to decay into a lighter real scalar singlet $S$ with mass $m_S=m_{H^0}/10$.}
\label{lambda5-vs-mH0-dixieme}
\end{figure}

Since the very minimal IDM scenario above  cannot account  for both the relic density and direct detection constraints at the same time, one question one must ask is whether 
this simple scenario 
of an IDM asymmetry creation could not be nevertheless at the origin of the  DM relic density  today in a simple way. This could in fact happen if the DM is made of a lighter specie, 
whose relic density would be due to the reprocessing of the inert doublet asymmetry into this specie. Such a reprocessing could for instance take place through decay.
For the scalar scenario we consider, this could be the case if there exists a lighter $Z_2$ odd particle, ``$S$", to  which the inert doublet states can decay. 
In this case, if the asymmetry is fully reprocessed into  this  lighter particle $S$, so that each inert scalar component gives one $S$ particle, the results of Figs.~\ref{sigmaH2-vs-mass} and \ref{lambda5-vs-mH0} are still fully valid provided the mass of the $S$ particle, $m_S$, is close to $m_{H_2}$. If instead it is sizably smaller, this requires to create more inert particles by a factor $m_{H_2}/m_S$. As an example, Figs.~\ref{sigmaH2-vs-mass-bis} and~\ref{lambda5-vs-mH0-dixieme} show the value of parameters we need to get the observed relic density for a ratio $m_{H_2}/m_S$ equal to 10. Note that such large $H_2$ asymmetry  cases, beside allowing smaller DM masses, also give relaxed lower bounds on the $\lambda_{3,4}$ couplings (in order to suppress sufficiently the symmetric part). Sizably  smaller values of these couplings are possible, relaxing accordingly the Landau pole constraints.
In order to reprocess the inert doublet asymmetry into such a $S$ specie, various possibilities could be considered. 

A simple possibility is to consider  $S$ as a $Z_2$-odd scalar singlet, into which the scalar  doublet  states decay sufficiently slowly to happen after the freeze-out of this singlet DM particle.
Such a decay can be accounted for by a ${\cal L}\ni \mu_S H_1^\dagger H_2 S$ renormalizable   interaction. If so, the main constraint to satisfy along such a scenario is, in order that the $S$ relic density  is  mainly produced from the IDM asymmetry, that the $S$ particles has a  $S^\dagger S\to SM\, SM$ annihilation channel  with  a large enough  cross section 
to leave 
a relic density smaller than the observed one at $S$ freeze-out. These annihilations can be accounted for by a ${\cal L}\ni \lambda_S  H_1^\dagger  H_1 S^\dagger S$  interaction.

\begin{figure}
\centering
    \includegraphics[scale=.68]{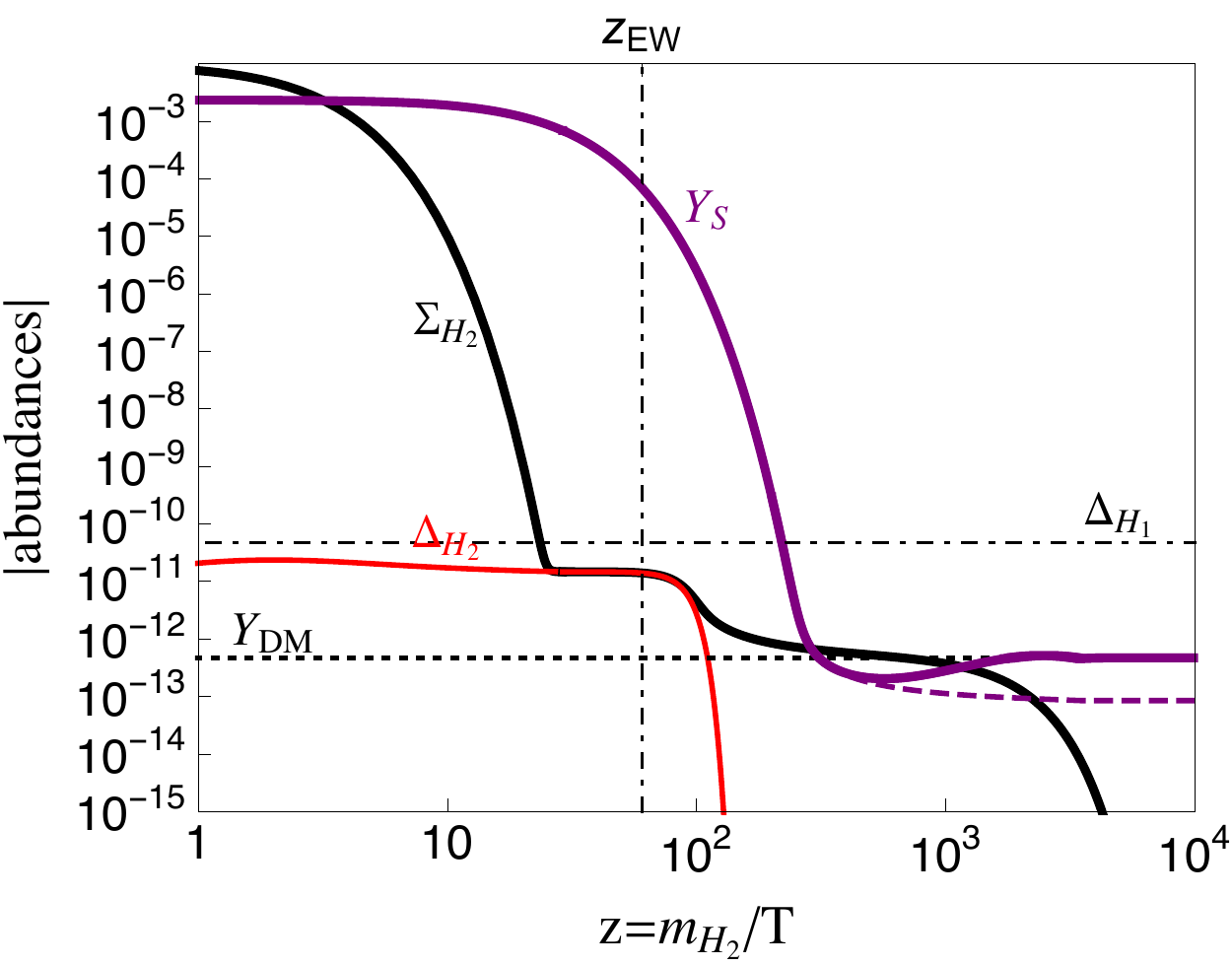}   
  \caption{Evolution of the various abundances as a function of $z=m_{H_2}/T$ in the case where the inert doublet decays into a real state $S$ after $S$ freezeout. Such an evolution has been obtained fixing the  $H_2$-related parameters to
  $m_{H_2}=10$~TeV, $\langle\sigma_{eff} v\rangle=0.5$~pb and $\delta m_{H_2} = 3\cdot 10^{-6}$~GeV (corresponding to $\lambda_5\approx   10^{-6}$),  the  $S$-related parameters to $m_S = 1$~TeV and $\langle\sigma_S v\rangle=4$~pb,   and the connector parameter controlling the decay rate to $\mu_S=5\cdot 10^{-6}$~GeV. The dashed purple curve shows what would be the evolution of the $S$ density if there were no $H_2\to H_1S$ decay.}
\label{Evolution-abundances-real-S-2}
\end{figure}

As an  example, if we take a real scalar singlet, and fix the parameters to be $m_S\sim 2$~TeV (400 GeV) and $m_{H_2}=10$~TeV, both conditions are fulfilled for $\lambda_S\gtrsim 0.6(0.1)$ and $\mu_S\lesssim4\cdot 10^{-5}$~GeV ($7\cdot 10^{-6}$~GeV). 
Also, the $\lambda_S$ interaction induces elastic scattering on nucleon through SM scalar exchange,  $\sigma_N=\lambda_S^2 m_N^4f_N^2/(\pi m_h^4m_S^2)$, where $m_N$ is the nucleon mass and the nucleon form factor is approximately given by $f_N\approx0.3$.
The LUX experiment constraint \cite{Akerib:2013tjd}, which for  $m_{DM}\gtrsim 100$~GeV is $\sigma_N\lesssim 1.2\cdot10^{-11}(m_S/\text{1 GeV})$~pb, is satisfied if $\lambda_S\lesssim 1.6\cdot 10^{-5} (m_S/\text{1 GeV})^{3/2}$. Combining both these lower and upper bounds on $\lambda_S$ leads to the lower bound $m_S\gtrsim300$~GeV~\cite{Cline:2013gha}.
In Fig.~\ref{Evolution-abundances-real-S-2} we show the evolution of the asymmetries we get as a function of the temperature for an example of parameter set which leads to the observed relic density.

 \begin{figure}
\centering
    \includegraphics[scale=.68]{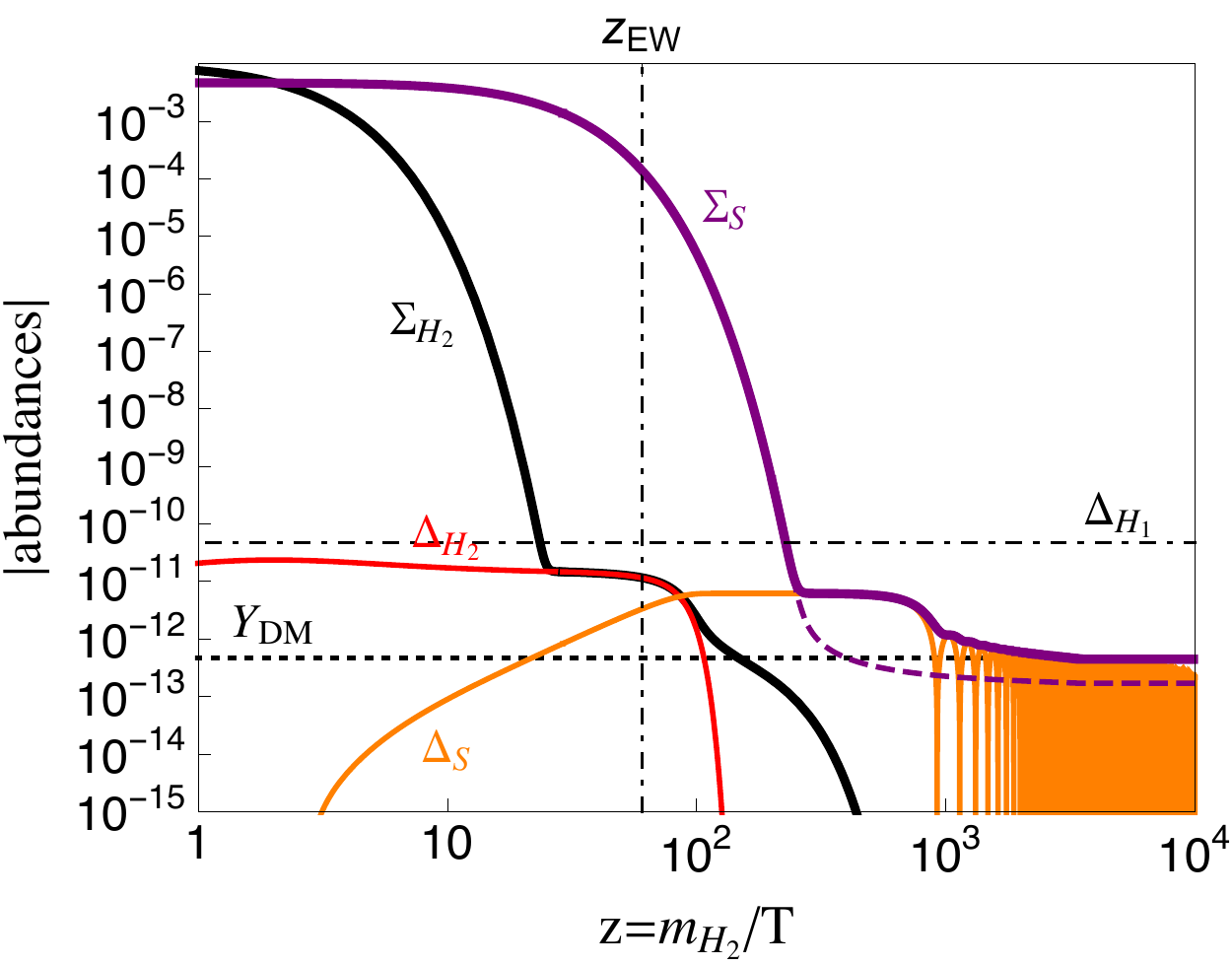}   
  \caption{Evolution of the various abundances as a function of $z=m_{H_2}/T$ in the case where the inert doublet decays into a complex state $S$ before $S$ freezeout, for  $m_{H_2}=10$~TeV, $\langle\sigma_{eff} v\rangle=1$~pb, $\delta m_{H_2} = 3\cdot 10^{-6}$~GeV (corresponding to $\lambda_5= 10^{-6}$), $m_S = 1$~TeV, $\langle\sigma_S v\rangle=4$~pb, $\delta m_S = 10^{-7}$~eV,  and $\mu_S=5\cdot 10^{-5}$~GeV. The dashed purple curve shows what would be the evolution of the $S$ density if there were no $H_2\to H_1S$ decay.}
\label{Evolution-abundances-complex-S-2}
\end{figure} 

Similarly to the fermion scenario considered in Ref.~\cite{Servant:2013uwa}, another  possibility is to consider instead that the decays occur when the freeze-out of the singlet particle $S$ has still not taken place. In this case, the inert doublet asymmetry  could also be at the origin of the DM relic density, if the singlet is a complex field and if the inert doublet asymmetry is reprocessed into a $S$ asymmetry.
Since inert doublet oscillations start at $T_{EW}$, this requires the reprocessing to be done prior to EWSB.
Imposing in addition for simplicity that the decay occurs after the $\lambda_5$ interaction has decoupled at $z_{\lambda_5}$, for example for $m_{H_2}=10$~TeV and $m_S\sim 2$~TeV, one needs $10^{-5}~\hbox{GeV}\lesssim \mu_S \lesssim 10^{-3}\,\hbox{GeV}$.
For this scenario to work, one has to make sure that the $S$ asymmetry created in this way is not washed-out by possible $S$-$S^\dagger$ oscillations. This requires that   terms as $\lambda'_S H_1^\dagger H_1 (S^2+h.c.)$ or $m_S^{\prime 2} S^2+h.c.$ are  sufficiently suppressed for the oscillations not to occur before $S$ freeze-out.  This means  the $S$ mass splitting $\delta m_S= (m_S^{\prime 2}+\theta(T_{EW}-T)\lambda'_S v^2)/2m_S$  must be  smaller than 
\begin{equation}
\delta m_S \lesssim 10^{-2}\cdot(z^S_{fo})^{-5/2}\cdot\left(\frac{m_S}{\text{1 TeV}}\right)^2\cdot\sqrt{\frac{\langle\sigma_Sv\rangle}{\text{1 pb}}} \text{ eV} \ ,
\end{equation}
with $z^S_{fo}\gtrsim 20$ the value of $m_S/T$ at which the $S$ freeze-out occurs, and $\langle\sigma_Sv\rangle$ the $S$ annihilation cross section.
Note that at temperature lower than $T^S_{fo}$, when the $S$ oscillations starts at $z^S_{osc}$, they can allow the $S$ annihilation to restart in the same way as for the inert doublet above.  Similarly to Eq.~(\ref{eq:DM-abundance-high-T-sym}), this causes a suppression of the $S$ asymmetry by a factor equal to $(1+\kappa_S \Sigma_S(z_{fo}^S)^{-1}$ with $\kappa_S=\langle \sigma_S v\rangle s{z}/H(z)z^S_{osc}$.
In Fig.~\ref{Evolution-abundances-complex-S-2} we show an example of evolution of the $H_2$ and $S$ asymmetries along such a scenario.

\section{Summary} 

In summary, if there exists an inert scalar doublet $H_2$, unless the $\lambda_5$ interaction is tiny, the inert doublet components will automatically develop an asymmetry from thermalization with the ordinary SM scalar doublet and lepton asymmetries. 
We have studied in details what is the fate of such an asymmetry at temperature below the value of the inert doublet scalar mass, $m_{H_2}$.
Beside being responsible for the asymmetry creation, the $\lambda_5$ interaction also controls the neutral component mass splitting (hence the $Z$-exchange direct detection rate) and induces a ``$k$-factor" suppression of the inert doublet asymmetry at temperature below $m_{H_2}$. On top of this suppression one can also have an extra  ``$\kappa$-factor" suppression, from the combined effect of DM oscillation (also induced by the $\lambda_5$ interaction) and DM symmetric annihilation. This leads to a scenario which chronologically occurs as represented in Fig.~\ref{sketch}. We showed that in the few-TeV range, there is an all region of parameter space where the DM asymmetry survives enough to lead to the observed DM relic density, Fig.~\ref{sigmaH2-vs-mass}, but this region turns out to lead to a too large $Z$-exchange direct detection contribution. As a result this scenario is nothing but excluded.

Next we looked at the possibility that the inert scalar asymmetry produced could still be at the origin of the observed DM relic density, which could be the case if it is reprocessed to a lighter specie, $S$, which satisfies the direct detection constraints.
We considered 2 scenarios where DM is made of a singlet odd under the $Z_2$ symmetry. a) Slow decay of the asymmetry into the (real or complex) singlet particle, occurring after $S$ freeze-out. b) Reprocessing of the inert scalar doublet asymmetry into a $S$ asymmetry through faster decays occurring before $S$ freeze-out. Both possibilities can lead to the observed relic density   provided the interaction causing the decay is small enough to induce this decay at the right time.

As most asymmetric DM scenarios, the framework we consider does not explain why the baryon and DM abundances are so similar.
Our scenario trades this abundance coincidence for a coincidence between the mass of the proton, the mass of the inert states, the mass of the dark matter particle, and the values of various couplings. Even if both abundances have same origin, these parameters must "cooperate"  to lead to a DM abundance so close to the baryon one.
Rather than providing a real explanation for the abundance coincidence, this scenario shows instead that 
the origin of the DM relic density could be of asymmetric origin, due to the generation of an inert scalar asymmetry related to the generation of a $B-L$ asymmetry at high temperature.

\vspace{5mm}
\section*{Acknowledgement}
\vspace{-3mm}
We acknowledge useful discussions with M.~Cirelli, G.~Servant, S. Tulin and M. Tytgat.
This work is supported by the FNRS-FRS, the IISN, the FRIA and an ULB-ARC and the Belgian Science Policy, IAP VI-11.

\vspace{1cm}
\noindent

\section*{Appendix}

\subsection*{Rates and cross sections}

 In Eq.~(\ref{boltzmann-eq}), the reaction density of the $\lambda_5$ scatterings for the $\eta^+$ (and similarly for $\eta^0$) is given by 
\begin{equation}
\gamma_{\lambda_5}=  \gamma^{\phi\phi}_{\eta\eta}+ \gamma^{\phi\eta}_{\phi\eta}  \ ,
\end{equation}  
where 
 \begin{equation}
\gamma^{ab}_{cd}=\frac{m_{H_2}^4}{64\pi\, z}\int_{4}^{\infty} dx\, \sqrt{x}\, K_1(z\sqrt{x})\,  \hat \sigma\left(ab\to cd\right)  \ .
\end{equation} 
with $\hat \sigma (ab\to cd )$ the reduced cross section. These are given by
\begin{align}
 \hat\sigma^s_{\lambda_5} (\phi\phi\to \eta\eta) & =\frac{3 \lambda_5^2}{2\pi }\sqrt{1-\frac{4}{x}} \, , \\
  \hat\sigma^t_{\lambda_5}(\phi\eta\to \phi\eta) &  =\frac{ 3\lambda_5^2}{2\pi }\left(1-\frac{1}{x}\right)^2 \, .
 \end{align} 
 In the non-relativistic limit, the corresponding rate is given by
 \begin{equation}
 \Gamma_{\lambda_5}\equiv\frac{ \gamma_{\lambda_5}}{n_{\eta^+}^{eq}}\equiv n_{\eta^+}^{eq}\langle  \sigma^s_{\lambda_5}v \rangle+n_{\phi^+}^{eq}\langle  \sigma^t_{\lambda_5}v \rangle \ , 
 \end{equation}
where 
\begin{align}
\langle  \sigma^s_{\lambda_5}v\rangle =\frac{3 \lambda_5 ^2}{32\pi m^2_{H_2} }  \, , \text{and}\quad
\langle  \sigma^t_{\lambda_5}v\rangle =\frac{3 \lambda_5 ^2}{16\pi m^2_{H_2} }  \, .
 \end{align}
In Eq.~(\ref{boltzmann-equation-zfo}) and~(\ref{Beq-naive-approach-final}), the effective cross section of the $H_2\bar H_2\to SM\ SM$ coannihilations    is given by \cite{Hambye:2009pw}
\begin{equation}\label{eff-cross-section-above-TEW}
\langle \sigma_{eff} v \rangle=\sum_{i,j}\langle\sigma_{ij}  v\rangle \frac{Y_i^{eq}}{\Sigma_{H_2}^{eq}}\frac{Y_j^{eq}}{\Sigma_{H_2}^{eq}} \simeq\frac{1}{64\, \pi\, m^2_{H_2}}\left(\frac{3}{8}g^4+\lambda_3^2+\lambda_4^2 \right)\ .
\end{equation}
where $g$ is the weak coupling constant. We neglected the $\lambda_5$ contribution, and 
the corrections due to the contributions proportional to $\langle v^2\rangle$.

 \

\subsection*{Analytical resolution of the Boltzmann equations}

The Boltzmann equations given in Eqs.~(\ref{Boltzmann-equation-1-1})-(\ref{Boltzmann-equation-1-3}) do not in general have a simple analytical solution. However, in the case of very fast oscillations, like it is the case here,  a good approximation consists in symmetrizing the equations  for $\Delta_{\eta^0}$ and $\Xi_{\eta^0}$,~i.e.~replacing Eqs.~(\ref{Boltzmann-equation-1-2})-(\ref{Boltzmann-equation-1-3}) by
\begin{align}\label{Boltzmann-equation-2}
	\frac{d\Delta_{\eta^0}}{dz}&=2 i\frac{ \delta m}{zH } \ \Xi_{\eta^0} -\frac{1}{2}\frac{\langle\sigma_{0 }  v\rangle s}{ zH}\  \Delta_{\eta^0}\ \Sigma_{\eta^0} \ ,\\\label{Boltzmann-equation-2-2}
	\frac{d\Xi_{\eta^0}}{dz}&=2 i\frac{  \delta m}{zH }\ \Delta_{\eta^0}-\frac{1}{2}\frac{\langle\sigma_{0 }  v\rangle s}{zH}\ \Xi_{\eta^0}\ \Sigma_{\eta^0}  \ .
\end{align}
In this approximation, the solutions for $\Delta_{\eta^0}$ and $\Xi_{\eta^0}$ are of the form 
\begin{align}\label{solution-delta-xi}
\Delta_{\eta^0}(z)&=f(z)\ \cos[g(z)] \ \ \text{,}\ \ \Xi_{\eta^0}(z)=i\ f(z)\ \sin[g(z)] \ .
\end{align}
Furthermore, since we are interested in oscillations happening after the freeze-out, we can neglect $\Sigma^{eq}_{\eta^0}$ in Eq.~(\ref{Boltzmann-equation-1-1}).
With these approximations, integrating from $z_{EW}$ to $z$ with the initial conditions $\Delta_{\eta_0}(z_{EW})=\Sigma_{\eta_0}(z_{EW})$ and $\Xi(z_{EW})=0$,  the analytical solutions of the Boltzmann equations    Eqs.~(\ref{Boltzmann-equation-1-1})-(\ref{Boltzmann-equation-2-2}) are given by Eq.~(\ref{solution-delta-xi})  and 
 \begin{equation}
 \Sigma_{\eta^0}(z)= \sqrt{\Delta^2_{\eta^0}(z)-\Xi^2_{\eta^0}(z)}=f(z) \ ,
 \end{equation}   
 with
\begin{align}\label{Boltzmann-equation-function} 
	f(z)&=\frac{\Sigma_{\eta_0}(z_{EW})}{  1+\frac{1}{2} \frac{\langle\sigma_{ 0}  v\rangle  \, s(z)}{H(z)}\left(\frac{z}{z_{EW}}-1\right)\Sigma_{\eta_0}(z_{EW})}\ , \\
	g(z)&=\frac{ \delta m}{H(z)}\left(\frac{z^2}{z^2_{EW}}-1\right)\ .
\end{align}
The abundance   $\Sigma_{\eta^0}$ decreases therefore  monotonically until it reaches an asymptotical value given by
\begin{equation}
\Sigma_{\eta^0}(z\gg z_{EW})=\frac{\Sigma_{\eta_0}(z_{EW})}{  1+\frac{1}{2} \frac{\langle\sigma_{ 0}  v\rangle  \, s(z)}{H(z)} \frac{z}{z_{EW}}\Sigma_{\eta_0}(z_{EW})}\ .
  \end{equation}
Note that despite appearance, the denominator doesn't depend on $z$, since   $sz/(Hz_{EW}) =12\sqrt{g_*}M_{Pl}T_{EW}/5\pi^2 $.

\end{document}